\documentclass[a4paper,11pt]{article}
\usepackage{jheppub} 
\usepackage{physics}
\usepackage{subfigure}
\usepackage{subfigure}
\usepackage{ascmac}
\usepackage{amsthm}
\usepackage{amsmath}
\usepackage{tikz}

\usepackage{pgfplots}
\pgfplotsset{compat=1.15}
\usepackage{mathrsfs}
\usetikzlibrary{arrows}

\allowdisplaybreaks

\newcommand{\F}[4]{
	F\left(
	\begin{array}{cc}
		#1, & #2 \\
		\multicolumn{2}{c}{#3}
	\end{array}; #4
	\right)
}

\preprint{KUNS-2971}

\title{
Photon sphere and quasinormal modes in AdS/CFT
}

\author{Koji Hashimoto, Kakeru Sugiura, Katsuyuki Sugiyama, Takuya Yoda}
\affiliation{Department of Physics, Kyoto University, Sakyo-ku, Kyoto 606-8502, Japan}

\emailAdd{koji@scphys.kyoto-u.ac.jp}
\emailAdd{sugiura@gauge.scphys.kyoto-u.ac.jp}
\emailAdd{sugiyama@scphys.kyoto-u.ac.jp}
\emailAdd{t.yoda@gauge.scphys.kyoto-u.ac.jp}

\abstract{
Photon spheres are the characteristic of general black holes, 
thus are a suitable touchstone for the emergence of gravitational spacetime 
in the AdS/CFT correspondence.
We provide a spectral analysis of an AdS Schwarzschild black hole near its photon sphere. 
We find that quasinormal modes near the photon sphere reflect the AdS boundary, 
resulting in a peculiar spectral pattern. 
Our large angular momentum analysis owes to an analogue to solvable Schr\"odinger equations 
such as an inverted harmonic oscillator and the P\"oschl-Teller model, 
with a Dirichlet boundary condition.
Through the AdS/CFT dictionary, it predicts the existence of a peculiar subsector in the large angular momentum spectrum of thermal holographic CFTs on a sphere.
}

\begin{document} 
\maketitle
\flushbottom

\section{Introduction}

Since the whole mechanism of the holographic principle has not yet uncovered, any possible indicator of the emergence of a gravitational spacetime is welcome to be tested in the dictionary of the AdS/CFT correspondence \cite{Maldacena:1997re}. 
As is popularly known, {\it photon spheres} are the characteristic of general black holes, 
needless to speak about the recent observation of an Einstein ring in the M87 black hole \cite{EventHorizonTelescope:2019dse}.
Therefore it is natural to study the photon spheres in the context of the AdS/CFT, to find whether
the photon spheres can be an indicator of the spacetime emergence, from the viewpoint of the dual CFT.

The photon spheres are best-undrstood in a particle picture as its eternal motion circulating the black hole. In general,
for spherical black holes, the photon sphere is a sphere on which photons can circulate the black hole forever. The orbit is unstable: the circulating photon can be swallowed by the black hole or escape from it, once perturbed infinitesimally. The effective motion is dictated by the null geodesics  determined by the effective potential which has a peak at the photon sphere. This potential hill
is made just by the combination of the gravitational attraction and the centrifugal force of the angular momentum of the photon.

The fundamental relation in the AdS/CFT dictionary 
is the coincidence of the spectra in the gravity side and the CFT side.
In the gravity side, the spectrum is nothing but the quasinormal modes (QNMs): the fluctuation spectrum
of (photon) fields in the background of a black hole. 
The field, once quantized, should correspond to a particle orbiting the black hole. Therefore
in general the spectra of the QNMs should also be affected by the photon sphere.

For our study of the photon sphere in the AdS/CFT, we find three important streams of research development.
First, recently, it was shown that a symmetry group $SL(2;\mathbf{R})$ dictates the QNM spectra near the photon sphere/ring of Schwarzschild and Kerr black holes in an asymptotically flat spacetime \cite{Raffaelli:2021gzh,Hadar:2022xag}. 
The analysis was further generalized to warped geometries \cite{Kapec:2022dvc,Chen:2023zvd} and related properties were studied \cite{Chen:2022fpl,Fransen:2023eqj}. 
On the other hand, the QNMs in the AdS Schwarzschild black hole has a long history:
they were first studied in \cite{Chan:1996yk,Horowitz:1999jd,KalyanaRama:1999zj}, and the general study in the WKB approximation was performed by Festuccia and Liu in \cite{Festuccia:2008zx} (see also \cite{Cardoso:2001bb,Moss:2001ga,Starinets:2002br,Konoplya:2002zu,Michalogiorgakis:2006jc,Siopsis:2007wn,Miranda:2008vb,Cardoso:2008bp,Daghigh:2008jz,Miranda:2009uw,Berti:2009wx,Denef:2009kn,Morgan:2009pn,Morgan:2009vg,Daghigh:2009fy,Gannot:2012pb,Aragon:2020tvq,Daghigh:2022uws,Fortuna:2022fdd} for various QNM analyses in AdS black holes, and see \cite{Berti:2009kk,Konoplya:2011qq} for reviews).
Thirdly, the direct consequence of the existence of the photon sphere is the Einstein ring in images of general black holes, and an imaging transform on holographic CFTs indeed was shown to produce such Einstein rings
\cite{Hashimoto:2018okj,Hashimoto:2019jmw}. 

In this paper, motivated by these works, we study QNMs near the photon sphere in the 
AdS Schwarzschild black hole. 
We find a peculiar form of the spectra associated with the photon sphere. 
Through the AdS/CFT, this shows the existence of a peculiar subsector in the large angular momentum spectrum of thermal holographic CFTs on a sphere at high temperature.


We first look at the existence condition of the photon sphere in the AdS Schwarzschild geometry. In fact, there exists a minimum angular momentum for the existence of the potential hill \cite{Festuccia:2008zx}. 
We estimate the minimum to find typical values of the angular momentum which we focus on.

To obtain the QNM spectrum of our concern, we map the wave equation for the QNMs to a Schr\"odinger equation with 
a potential hill. The photon sphere is located at the top of the Schr\"odinger potential hill around which the potential can be approximated by an inverted harmonic oscillator. 
At a large angular momentum, the approximation is better, as expected from the results reported in the asymptotically flat case \cite{Hadar:2022xag}. 
We find that a large angular momentum limit brings the quantum mechanics to a simple form: a potential hill at the photon sphere, and a
Dirichlet boundary at the asymptotic AdS infinity. We obtain an explicit relation between the CFT spectrum and the energy spectrum of the Schr\"odinger equation. At high temperature, we find an analytic expression for the Schr\"odinger potential. 

The Schr\"odinger equation is not integrable, thus to grasp a possible universal feature of the energy spectra, 
we resort to solvable quantum mechanics which share the feature of our Schr\"odinger equation. In fact, the solvable models have been used in the analyses of QNMs \cite{Ferrari:1984ozr}, and presently can provide an analytic expression for the energy spectrum. We use
an inverted harmonic oscillator and the P\"oschl-Teller
model, both of which have the potential hill. 

In solving those, we find that the existence of the AdS boundary is essential. Even when we bring the AdS boundary to the spatial infinity effectively (which is possible by taking the large angular momentum limit), the effect of the boundary still remains crucial, and the spectrum does not lead to that of the asymptotically flat case.

As these two solvable models provide a spectral pattern similar to each other, we claim that our system of the QNMs near the photon sphere in the AdS Schwarzschild black hole should 
share the pattern. Through the AdS/CFT dictionary, the obtained spectrum corresponds to 
the spectrum of a peculiar subsector of the thermal CFT on a sphere.

The organization of this paper is as follows.
In Sec.~\ref{sec:2}, we examine the existence condition of the photon sphere in the AdS Schwarzschild. Then in 
Sec.~\ref{sec:3}, we relate the QNM equation with a Schr\"odinger equation of a quantum mechanics with a potential hill, and find an explicit relation between the QNM spectrum at a large angular momentum and the quantum mechanical energy. In Sec.~\ref{sec:4}, we solve the spectrum of solvable quantum models analogous to the one given in Sec.~\ref{sec:3}, and obtain the pattern of the QNM spectrum.
Sec.~\ref{sec:5} is for our summary and discussions. In App.~\ref{sec:A} we summarize the representation theory of $SL(2;\mathbf{R})$, to illustrate that it is broken by the AdS boundary. App.~\ref{sec:B} is for detailed information on the solvable models used in Sec.~\ref{sec:4}.

\section{Photon sphere in AdS black hole}
\label{sec:2}
In this section, the photon sphere in the AdS Schwarzschild black hole spacetime is described.
We demonstrate that a test field feels a photon sphere for any Hawking temperature $T$, taking a sufficiently large angular momentum $l>l_\mathrm{min}$, where $l_\mathrm{min}$ is a $T$-dependent minimum.
Then, the $T$-dependence of $l_\mathrm{min}$ is estimated.

In the cases in the ordinary Schwarzschild black hole, nonzero angular momentum always produces a photon sphere.
In the asymptotic AdS cases, however, the centrifugal potential is buried by the AdS curvature and the photon sphere disappears for small $l$.

\begin{figure}[t]
    \centering
    \includegraphics[height=5cm]{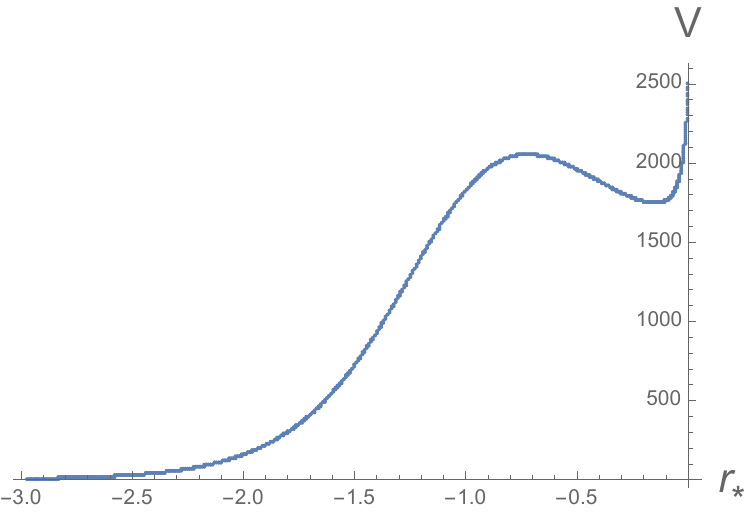}
    \caption{The effective potential $V$ in the tortoise coordinate $r_*$.
    The peaks near $r_*=-0.6$ and $r_*=0$ represent the photon sphere and the AdS boundary, respectively.
    For this figure we took $d=4$, $r_0/l_0=4\sqrt{3}/9$, and $l=40$. 
    }
    \label{fig: V in r*}
\end{figure}

Let us consider a scalar field $\Phi$ with mass $\mu$ in the $d$-dimensional AdS Schwarzschild spacetime 
\begin{align}
    ds^2=-f(r)dt^2+\frac{dr^2}{f(r)}+r^2d\Omega_{d-2}^2,\qquad f(r)=1+\frac{r^2}{l_0^2}-\qty(\frac{r_0}{r})^{d-3},
\end{align}
where $d\geq4$, $l_0$ is the AdS radius, and $r_0$ corresponds to the black hole temperature $T$.\footnote{
The horizon radius $r_\mathrm{h}$ is related to this parameter $r_0$ as $r_0 = r_\mathrm{h}(1 + r_\mathrm{h}^2/l_0^2)^{1/(d-3)}$.
This is a monotonic function of $r_\mathrm{h}$. The black hole temperature $T$, which is equal to the dual CFT temperature, is given by $T = ((d-1)(r_\mathrm{h}/l_0)^2 + (d-3))/4\pi r_\mathrm{h}$. At high temperature (large black holes, $r_\mathrm{h}\gg l_0$), we find $T\sim (d-1)r_\mathrm{h}/4\pi l_0^2$ and $r_0 \sim r_\mathrm{h}(r_\mathrm{h}/l_0)^{2/(d-3)}$, thus $r_0 \sim l_0 (4\pi l_0 T/(d-1))^{(d-1)/(d-3)}$. This means that the parameter $r_0$ is a monotonic function of $T$, at high temperature.
}
Decomposing the scalar field to the spherical harmonics as $\Phi=e^{-i\Omega t}Y_{lm}(\mathrm{angles})\psi(r)r^{1-d/2}$, the Klein-Gordon equation reduces to a Schr\"odinger-like equation \cite{Zerilli:1970wzz}
\begin{align}
    \qty(\dv[2]{r_*}+\Omega^2-V(r))\psi(r)=0,
    \label{Sch}
\end{align}
where 
\begin{align}
    r_* \equiv \int^r_\infty \frac{dr'}{f(r')}
\label{tort}
\end{align}
is the tortoise coordinate (the AdS boundary is at $r_*=0$) and $V$ is the effective potential defined by
\begin{align}
    V(r)=f\qty(\frac{(l+d/2-1)(l+d/2-2)}{r^2}+\qty(\frac{d}{2}-1)\qty(\frac{d}{2}-2)\frac{f-1}{r^2}+\qty(\frac{d}{2}-1)\frac{f'}{r}+\mu^2).
        \label{eq: Veff in SAdS}
\end{align}
The potential in the tortoise coordinate $r_*$ is illustrated in Fig.~\ref{fig: V in r*}.

The potential \eqref{eq: Veff in SAdS} is found to be a sum of the potential in the flat case and corrections to it:
\begin{align}
    \begin{aligned}
    V(r)=&\qty(1-\qty(\frac{r_0}{r})^{d-3})\qty(\frac{(l+d/2-1)(l+d/2-2)}{r^2}+\qty(\frac{d}{2}-1)^2\frac{r_0^{d-3}}{r^{d-1}}+\mu^2_\mathrm{eff})\\
    &+\frac{1}{l^2_0}\qty(\mu^2_\mathrm{eff}r^2+\qty(\frac{d}{2}-1)^2\qty(\frac{r_0}{r})^{d-3}+\qty(l+\frac{d}{2}-1)\qty(l+\frac{d}{2}-2)).
    \end{aligned}
\end{align}
The effect of the AdS curvature consists of 1) the effective mass $\mu^2_\mathrm{eff}=\mu^2+d(d-2)/4l_0^2$, 2) the $r^2$ term typical of AdS, and 3) the additional attractive term from the black hole; all of these tend to make the photon sphere vanish.

If $l$ is large enough, the effective potential \eqref{eq: Veff in SAdS} of the scalar field reduces to the potential of the massless geodesic
\begin{align}
    V(r)\approx f(r)\frac{l^2}{r^2}.
    \label{geodesicV}
\end{align}
Because the AdS effect here causes only the constant shift to the flat potential, the photon sphere always exists. Denoting the location of the photon sphere as $r=r_{\rm PS}$, we find the location in the large $l$ limit  
\begin{align}
    r_{\rm PS} \simeq \qty(\frac{d-1}{2})^{1/(d-3)}r_0=:\tilde{r}.
    \label{geops}
\end{align}
Therefore, the photon sphere must exist for any values of the parameters $d$ and $r_0$, taking a sufficiently large but finite $l$.

\begin{figure}[t]
    \centering
    \includegraphics[height=6cm]{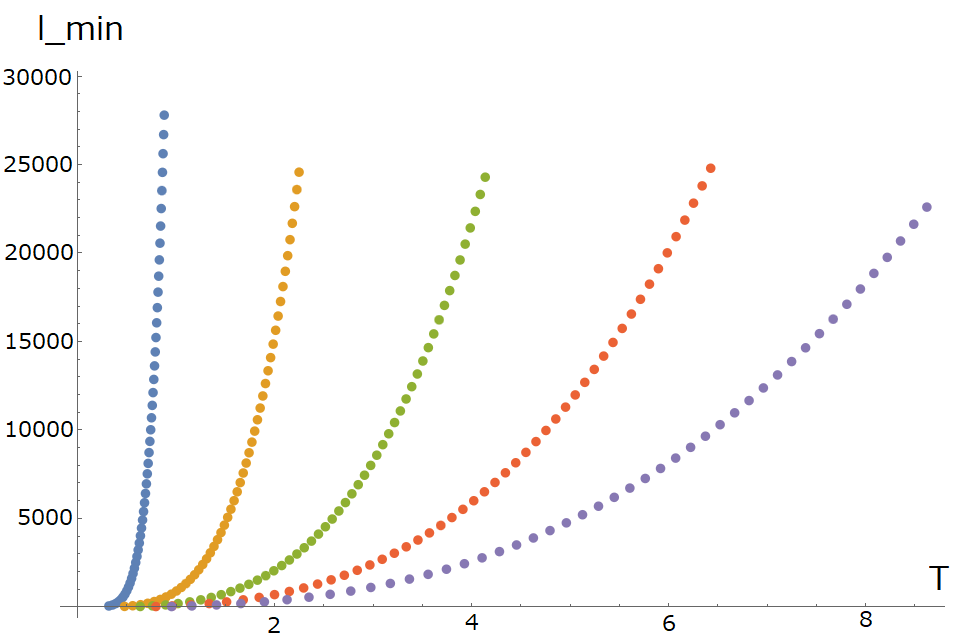}
    \caption{The $T$-dependence of the minimum angular momenta $l_\mathrm{min}$ to generate a photon sphere in various spacetime dimensions $d$. From the blue line to the purple, $d$ is 4 to 8. For this figure we took $\mu=0$.}
    \label{fig: lmin in ds}
\end{figure}

Indeed, we can numerically confirm that there is a minimum angular momentum $l_\mathrm{min}$ such that the photon sphere exists for any $l>l_\mathrm{min}$ (Fig. \ref{fig: lmin in ds}).
The value of $l_\mathrm{min}$ increases as the Hawking temperature $T$ does. Now we note that $T$ grows monotonically with respect to $r_0$ when $T$ is above the Hawking-Page transition temperature $T_c=(d-2)/2\pi l_0$, that is, $r_0>2^{1/(d-3)}l_0$.

\begin{figure}[t]
    \centering
    \includegraphics[height=4cm]{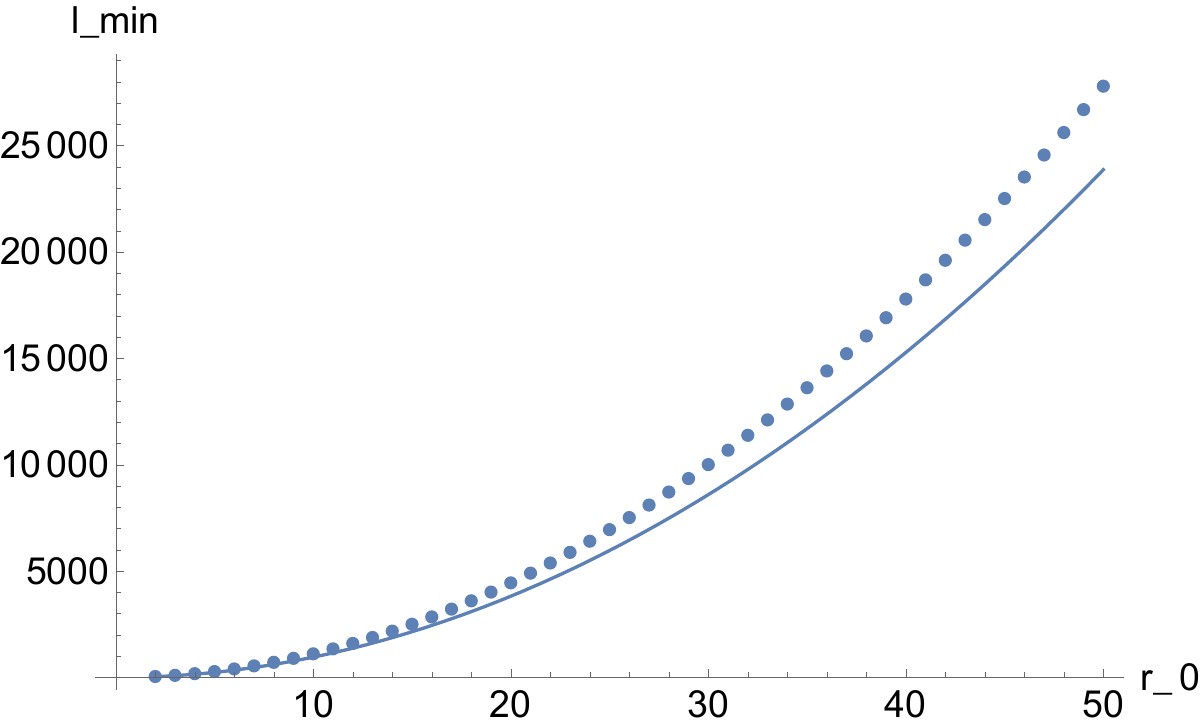}
    \includegraphics[height=4cm]{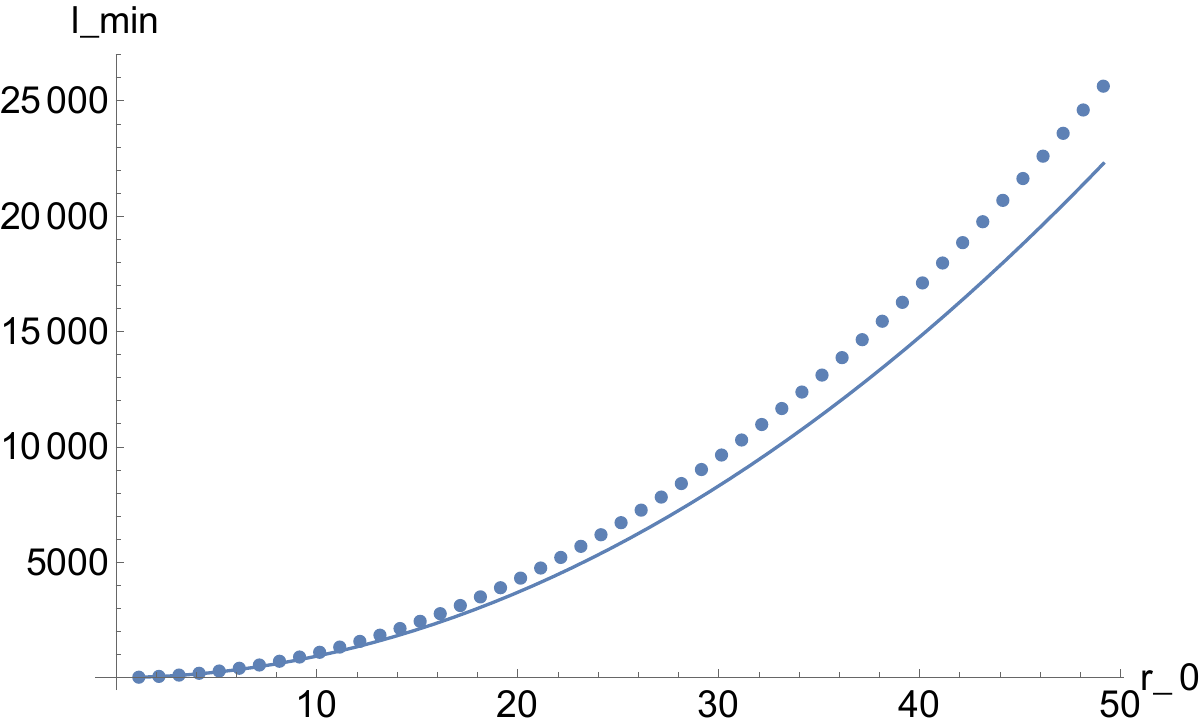}
    \caption{The $r_0$-dependence of $l_\mathrm{min}$ in $d=4$ (left panel) and $d=8$ (right panel).
    The dotted lines and the solid lines represent the numerical calculation and the analytic estimation \eqref{eq: lmin_estimate} respectively.}
    \label{fig: lmin_estimation}
\end{figure}

We can estimate the expression of $l_\mathrm{min}=l_\mathrm{min}(r_0)$ with an analytic calculation. In \cite{Festuccia:2008zx}, the critical value $l_\mathrm{min}$ of the angular momentum has been found as a solution of $V'(r_\mathrm{PS})=V''(r_\mathrm{PS})=0$ and obtained in a large $\mu$ limit. Now, instead, we provide the expression for $\mu=0$, at a large $l$.
It is expected that the location of the photon sphere $r_\mathrm{PS}$ approaches the geodesic one \eqref{geops} as $l$ increases. The difference can be estimated in the large $l$ expansion, as
\begin{align}
    r_\mathrm{PS} = 
    \tilde{r}
    -\frac{V'(\tilde{r})}{V''(\tilde{r})}+\order{\frac{1}{l^3}}.
    \label{rpsnext}
\end{align}
Regarding the condition that the photon sphere exists, which is expressed as $V''(r_\mathrm{PS})<0$, as the restriction for $l$, the value of $l$ is bounded from below by a bound given by the function of $r_0$.
If we naively use the expression of $V''$ and \eqref{rpsnext} and truncate them at the first subleading order in the large $l$ expansion, we find the following expression for the lower bounds, for example:
\begin{align}
    \begin{aligned}
        d=4:&\qquad l^2>\frac{729}{8}\qty(\frac{r_0}{l_0})^4+\frac{9}{2}\qty(\frac{r_0}{l_0})^2-\frac{4}{3},\\
        d=8:&\qquad l^2>\frac{78}{5}2^{\frac{1}{5}}7^{\frac{4}{5}}\qty(\frac{r_0}{l_0})^4+9\qty(\frac{2}{7})^{\frac{3}{5}}\qty(\frac{r_0}{l_0})^2-\frac{690}{49}.
    \end{aligned}
        \label{eq: lmin_estimate}
\end{align}
Fig.~\ref{fig: lmin_estimation} is the plot of these analytic bounds (solid lines) and the actual lower bound found numerically (dots). The qualitative behavior of $l_\mathrm{min}$ is well captured.\footnote{The difference between the analytic estimation and the numerical one grows as $r_0$ (or $T$) increases.
In the derivation of the bound \eqref{eq: lmin_estimate}, we adopt a naive truncation at the first sub-leading order, while 
the leading $\order{l^2}$ term and the sub-leading $\order{l^0}$ term are comparable to each other. It means that the expression \eqref{eq: lmin_estimate} is not compatible with the large $l$ expansion. We do not locate the reason why nevertheless \eqref{eq: lmin_estimate} looks a good approximation in Fig.~\ref{fig: lmin_estimation}.
}
From the analytic estimate,
we find that the minimum angular momentum $l_\mathrm{min}$ would grow as $\order{r_0^2}$, that is, as $\order{T^{2(d-1)/(d-3)}}$.

\section{From quasinormal modes to quantum mechanics}
\label{sec:3}

In this section we derive a quantum mechanical model which provides the quasinormal mode spectra, in the large angular momentum approximation.
The goal is, from the equation \eqref{Sch}, in the large $l$ approximation, 
to derive a potential problem of the form
\begin{align}
    \left(
	-\dv[2]{x} + \widetilde{V}(x)
    \right) \psi(x)
    = E\psi(x),
    \label{QM}
\end{align}
where the potential $\widetilde{V}(x)$ is approximated by 
$\widetilde{V}(x) \simeq - \frac{x^2}{4}$ around the top of the potential $x=0$.

\subsection{Potential hill relates to the quantum-mechanical model}

When the photon sphere exists, we can always expand the potential around the photon sphere. 
First, let us evaluate the potential $V(r)$ given in \eqref{eq: Veff in SAdS} around the top of the potential,
\begin{align}
    V(r) = V(r_\mathrm{PS}) + \frac12 (r-r_\mathrm{PS})^2 V''(r_\mathrm{PS}) + \cdots \, .
\end{align}
Note that here $V''$ is negative.
Then converting the coordinate $r$ to the tortoise coordinate $r_*$ near the top, and denote the difference 
from the top as $\delta r_*$ in the tortoise coordinate, we obtain
\begin{align}
    V = V(r_\mathrm{PS}) + \frac12 (\delta r_*)^2 V''(r_\mathrm{PS}) f(r_\mathrm{PS})^2 + \cdots \, .
\end{align}
Then we perform the following coordinate transformation
\begin{align}
    x \equiv \left(-2V''(r_\mathrm{PS}) f(r_\mathrm{PS})^2\right)^{1/4} \delta r_* .
    \label{def of x}
\end{align}
Together with the definition
\begin{align}
    E \equiv \frac1{\sqrt{-2V''(r_\mathrm{PS})} f(r_\mathrm{PS})}\left(\Omega^2 - V(r_\mathrm{PS})\right) ,
    \label{Eomega}
\end{align}
we arrive at the equation \eqref{QM} with 
\begin{align}
    \widetilde{V}(x) = - \frac{x^2}{4} + {\cal O} (x^3) .
\end{align}

Next, we shall evaluate the potential $\widetilde{V}(x)$ and the relation \eqref{Eomega} in various limits,
including the large $l$ expansion and the high temperature limit $r_0 \gg l_0$.

\subsection{Large \texorpdfstring{$l$}{TEXT} expansion}

\subsubsection{Quasinormal modes and quantum mechanical spectrum}

We can evaluate the coefficients in \eqref{Eomega}, in the large $l$ approximation. First, we redefine the frequency as
\begin{align}
    \Omega = \sqrt{V(r_\mathrm{PS})} + \widetilde{\Omega}.
    \label{ot}
\end{align}
At large $l$, this amounts to looking closely at the energy region near the photon sphere, since 
we regard $\widetilde{\Omega}$ to be
${\cal O}(l^0)$, that is, parametrically smaller compared to $\sqrt{V(r_\mathrm{PS})}$.
Using the definition \eqref{ot}, the relation \eqref{Eomega}
is rewritten as
\begin{align}
    \Omega = \sqrt{V(r_\mathrm{PS})} 
    + f(r_\mathrm{PS})\sqrt{\frac{-V''(r_\mathrm{PS}) }{2V(r_\mathrm{PS})}} E 
    - \frac{1}{2\sqrt{V(r_\mathrm{PS})}}\widetilde{\Omega}^2 \, .
\end{align}
In the large $l$ expansion, the first term is ${\cal O}(l)$, and the second term is ${\cal O}(l^0)$, while the third term is
${\cal O}(l^{-1})$. This means that, in the large $l$ expansion, we find
\begin{align}
    \Omega = a+b E 
    +  {\cal O}(l^{-1})\, ,
\end{align}
where 
\begin{align}
    a \equiv \sqrt{V(r_\mathrm{PS})} , \qquad
    b \equiv  f(r_\mathrm{PS})\sqrt{\frac{-V''(r_\mathrm{PS}) }{2V(r_\mathrm{PS})}} .
\end{align}
An explicit large $l$ expansion leads to
\begin{align}
a  
&= \sqrt{\left(\frac{1}{l_0^2} + \frac{1}{r_0^2}\frac{d-3}{d-1}\left(\frac{2}{d-1}\right)^{2/(d-3)}\right) 
\left(l+ \frac{d}{2}-1\right)\left(l+\frac{d}{2}-2\right)}+  {\cal O}(l^{-1}) \, , \\
    b
&    = \sqrt{(d-3)\left(\frac{1}{l_0^2} 
    + \frac{1}{r_0^2}\frac{d-3}{d-1}\left(\frac{2}{d-1}\right)^{2/(d-3)}\right)}+  {\cal O}(l^{-2}) \, .
\end{align}
So, in summary, we find a formula for the relation between the quasinormal mode spectra $\Omega$ and
the quantum mechanical energy $E$ of \eqref{QM} in the large $l$ approximation as
\begin{align}
    \Omega = 
    \sqrt{\left(\frac{1}{l_0^2} 
    + \frac{1}{r_0^2}\frac{d\! -\! 3}{d\! -\! 1}\left(\frac{2}{d\! -\! 1}\right)^{2/(d-3)}\right)}
    \left[\sqrt{\left(l\!+\! \frac{d}{2}\! -\! 1\right)\left(l\! +\! \frac{d}{2}\! -\! 2\right)} + \sqrt{d\!-\! 3} \, E\right]
    +  {\cal O}(l^{-1})\, .
\label{omegaE}
\end{align}
Once one solves the quantum mechanical model \eqref{QM}, substituting $E$ into the equation above derives the quasinormal mode spectra $\Omega$ in the large angular momentum approximation.

\subsubsection{On the validity of inverted harmonic oscillator}

While the energy spectra $E$ to be substituted into \eqref{omegaE} will be evaluated in the following section, 
we here study the shape of the potential $V(x)$. It is an inverted harmonic potential around the top of the potential, and we will see how good the approximation is at large $l$.

Near the photon sphere at $r=r_\mathrm{PS}$, the potential is expanded with the displacement $\delta r\ll r_\mathrm{PS}$ as
\begin{align}
    V(r_\mathrm{PS}+\delta r)=V(r_\mathrm{PS})+\frac{1}{2}V''(r_\mathrm{PS})(\delta r)^2+\frac{1}{3!}V^{(3)}(r_\mathrm{PS})(\delta r)^3+\cdots.
\end{align}
This is approximated by an inverted harmonic oscillator when $\delta r$ satisfies
\begin{align}
    \delta r\ll\frac{3V''(r_\mathrm{PS})}{V^{(3)}(r_\mathrm{PS})}.
\end{align}
This means that the approximation by the iniverted harmonic oscillator is valid in the range
\begin{align}
    V(r_\mathrm{PS})\geq V(r)\gg V(r_\mathrm{PS})+\frac{9}{2}\frac{(V''(r_\mathrm{PS}))^3}{(V^{(3)}(r_\mathrm{PS}))^2} \, .
\end{align}
The potential range $\Delta V=|9(V''(r_\mathrm{PS}))^3/2(V^{(3)}(r_\mathrm{PS}))^2|$ is of $\order{l^2}$, and the range is wide enough for a large $l$. Therefore, as long as the energy $\Omega^2$ of the quasinormal mode is in this range, one expects a safe use of the inverted harmonic potential. 
However, as we will discuss below, we argue that the spectra are not determined solely by the potential shape near the top.

\subsubsection{AdS boundary is a hard wall}

Let us turn to the shape of the potential near the AdS boundary.
We will show that the AdS boundary can be regarded as a hard-wall potential, where a Dirichlet condition is imposed on the field.

Near the boundary $r \sim \infty$, the tortoise coordinate \eqref{tort} is well-approximated as
\begin{align}
    r_* = -\frac{l_0^2}{r} ,
    \label{tort1}
\end{align}
because the effect of the black hole disappears and the geometry is almost equal to that of the pure AdS.
Using this expression, the potential $V(r)$ in \eqref{eq: Veff in SAdS} is approximated as
\begin{align}
    V \simeq
    \frac{1}{l_0^2}
    \left(l+\frac{d}{2}-1\right)
 \left(l+\frac{d}{2}-2\right)
 + \frac{1}{r_*^2}
     \left[\frac{d}{2}\left(\frac{d}{2}-1\right)+\mu^2l_0^2 \right].
     \label{wall}
\end{align}
The second term diverges at the AdS boundary $r_*=0$. So, it works as a wall by which any wave is bounced back.

The reason why this is a hard wall is as follows. In the large $l$ limit, the first term in \eqref{wall} is pushed up as a constant but being very high. The effect of the second term appears only when these two terms are comparable, 
\begin{align}
    r_* \sim l_0 
    \left[\frac{d}{2}\left(\frac{d}{2}-1\right)+\mu^2l_0^2 \right]^{1/2} \frac{1}{l},
\end{align}
at a large $l$. This point is very close to the AdS boundary $r_*=0$, and in the large $l$ limit, it coincides with the AdS boundary. Therefore, the AdS boundary behaves as a hard wall at a large $l$.

The existence of the hard wall is crucial in the analysis, as we will find in the next section. In fact, the difference between the asymptotically flat case and the AdS case was already noted in \cite{Cardoso:2008bp}.

\subsection{Analytic potential at high temperature}

In the above, we saw that the potential is approximated by the inverted harmonic oscillator around the top, and there exists a hard wall at the AdS boundary. In the next section we evaluate $E$ in systems which possess these characteristics.
In this subsection, we consider the case of a high temperature and will find an analytic shape of the potential $\widetilde{V}(x)$ which actually possesses the features described above.\footnote{Note that we take the limit $l\to\infty$ first and then a high temperature limit, otherwise the potential hill would disappear as we described in Sec.~\ref{sec:2}.}

First, we evaluate the tortoise coordinate \eqref{tort} as 
\begin{align}
    r_* = \int^r_\infty dr'\left[1 + \frac{(r')^2}{l_0^2} - \frac{r_0^{d-3}}{(r')^{d-3}}\right]^{-1}
    = r_0 \int^{r/r_0}_\infty 
    ds\left[1 -\frac{1}{s^{d-3}}+ \frac{r_0^2}{l_0^2}s^2 \right]^{-1} ,
\end{align}
where we made a change of variable $s\equiv r/r_0$. It is understood that in the high temperature case
$r_0 \gg l_0$, we find an approximation
\begin{align}
    r_* \simeq r_0 \int^{r/r_0}_\infty 
    ds\left[ \frac{r_0^2}{l_0^2}s^2 \right]^{-1} = -\frac{l_0^2}{r}
    \label{tort2}
\end{align}
as long as 
\begin{align}
    r \gg 
r_0^{\frac{d-3}{d-1}} l_0^\frac{2}{d-1}.
\end{align}
Note that the latter condition is satisfied for the photon sphere $r=r_\mathrm{PS}$ given in \eqref{geops} at large $l$, under the high temperature condition $r_0 \gg l_0$. Thus, although \eqref{tort2} is exactly the same as the near-boundary expression \eqref{tort1}, we can use it even near the top of the potential, at high temperature.

Using \eqref{tort2}, the large $l$ potential \eqref{geodesicV} is written as
\begin{align}
    V & \simeq
    l^2
 \left[
     \frac{1}{l_0^2}+\frac{1}{l_0^4}r_*^2 - \frac{r_0^{d-3}}{l_0^{2(d-1)}}(-r_*)^{d-1}
 \right]
 & 
 \left(r_*<0, \, |r_*| \ll r_0^{\frac{3-d}{d-1}} l_0^{\frac{2(d-2)}{d-1}}
 \right)
 \label{Vasymp}\\
 V & = \infty &  (r_* \geq 0)
\end{align}
The latter is to make sure that there exists a hard wall at $r_*=0$, as explained in the previous subsection.
The potential is the same as the one found in \cite{Festuccia:2008zx} at a WKB approximation.

With the linear redefinition \eqref{def of x} of $r_*$, we can rewrite this potential into the form \eqref{QM}. 
A simple calculation leads to the quantum mechanical potential $\widetilde{V}(x)$
whose hill top is located at $x=0$ with the hard wall located at $x=L$,
\begin{align}
    \widetilde{V}(x) = \frac{1}{4(d-3)}
    \left[
    (L-x)^2 - \beta^{d-3}(L-x)^{d-1} -\left(L^2-\beta^{d-3}L^{d-1}\right)
    \right],
    \label{Vuni}
\end{align}
where
\begin{align}
    \beta \equiv (d-3)^{-1/4} 2^{-1/2} \frac{1}{\sqrt{l}} \frac{r_0}{l_0}, \quad
    L \equiv \beta^{-1} \left(\frac{2}{d-1}\right)^{\frac{1}{d-3}}.
    \label{defL}
\end{align}
Note that in these expressions we keep only leading terms at the high temperature $r_0 \gg l_0$ and also the large $l$, and
we have adjusted the constant term of the potential
so that the height of the potential top is tuned to $\widetilde{V}=0$, for our later purpose.

\begin{figure}[t]
    \centering
    \includegraphics[height=5cm]{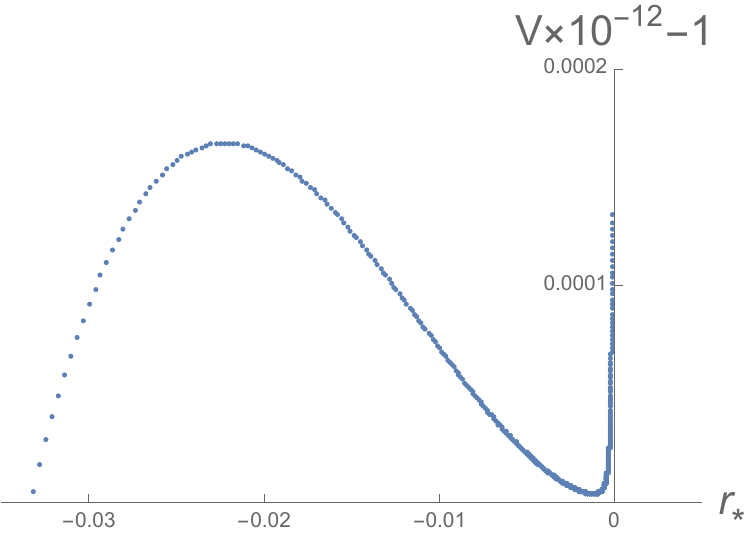}
    \caption{The effective potential $V$ in the tortoise coordinate $r_*$, at high temperature.
    For this figure we took $d=4$, $\mu=0$, $r_0=30$, $l=10^6$, in the unit $l_0=1$. 
    }
    \label{fig: V in r* at high temperature}
\end{figure}

To illustrate this potential, we consider the case $d=4$. We obtain
\begin{align}
    \widetilde{V}(x) = -\frac14 \, x^2 + 2^{-5/2} \frac{1}{\sqrt{l}} \frac{r_0}{l_0} \, x^3 ,
\end{align}
and the wall position is at
\begin{align}
    L = \frac{2^{3/2}}{3} \sqrt{l}\frac{l_0}{r_0} .
    \label{wall_loc}
\end{align}
The system is defined in the region $x<L$, and all waves are reflected at the AdS boundary hard wall $x=L$. The potential approximation is not trusted for large negative values of $x$.
In Fig.~\ref{fig: V in r* at high temperature}, we plot the potential \eqref{eq: Veff in SAdS} numerically evaluated at high temperature $r_0/l_0=30\gg 1$. We can confirm the potential shape \eqref{Vuni}, with the hard wall at the AdS boundary.

How far is this wall from the top of the hill? The existence condition of the photon sphere \eqref{eq: lmin_estimate} shows, at high temperature $r_0/l_0 \gg 1$, that we have the condition $l^2 \gg {\cal O}((r_0/l_0)^4)$. The wall location \eqref{wall_loc} then implies $L\gg {\cal O}(1)$. We conclude that the wall is located very far from the potential hill, in the metric of this $x$ space.
In addition, note that near the wall the potential approaches a new bottom, $V'(L)=0$, which means that
the whole potential cannot exactly be equal to the inverted harmonic potential for any choice of the wall position.

\section{Solving quasinormal modes respecting AdS boundary}
\label{sec:4}

In this section,
we analyze the quasinormal mode spectrum by solving quantum mechanical models, through the relation studied in Sec.~\ref{sec:3}.
The potential of the quantum mechanical model \eqref{QM} is obtained only in numerical manner, and it is numerically demanding to calculate the spectrum at the large $l$, as mentioned in \cite{Horowitz:1999jd}.
Thus instead, we use solvable models, the inverted harmonic oscillator and the P\"{o}schl-Teller model,  which share the same shape of the potential as that of \eqref{QM}. 
We show that the discrete spectrum is located on a peculiar curve in the complex energy $E$-plane, reflecting the potential hill of a photon sphere and the AdS boundary.
The effect of the boundary does not vanish even when the boundary is sent to the spatial infinity since decaying modes $\Im E<0$ grow near the boundary.
The solvable models result in the 
unique pattern of the spectrum, and in particular, the curve is rather insensitive to the shape of the potential tail. This universality leads us to state that the pattern of the QNM spectrum is captured by that of
the solvable models.

\subsection{Inverted harmonic oscillator}

Recall that
the Kelin-Gordon equation on the AdS Schwarzschild background
reduces to a quantum mechanical model \eqref{QM}.
The potential has a hill $\widetilde{V}(x)\simeq -\frac{x^2}{4}$ whose top corresponds to the photon sphere.
The energy $E$ is measured from the top of the hill.

In this subsection we replace the potential with the exact inverted harmonic potential
\begin{align}
    \widetilde{V}(x) \rightarrow -\frac{x^2}{4},
\end{align}
and solve the Schr\"{o}dinger equation
\begin{align}
    \left(
	-\dv[2]{x} - \frac{x^2}{4}
    \right) \psi(x)
    = E\psi(x).
\end{align}
We impose the Dirichlet boundary condition,
\begin{align}
    \psi(L) = 0,
    \quad L \gg 1.
\end{align}
Also we assume that
the wave function $\psi(x)$ has only the out-going mode in a region $x\rightarrow-\infty$ so that the amplitude decays with respect to time.
The potential hill at $x=0$ represents the photon sphere, while the Dirichlet boundary at $x=L$ represents the AdS boundary.
The out-going mode in the region $x\rightarrow-\infty$ corresponds to a scalar mode that is absorbed by the black hole: the horizon condition.

The Schr\"{o}dinger equation is solvable\footnote{
See App.~\ref{sec:B-1} for more detailed discussion on the spectrum.
}, and 
its solution is called the parabolic cylinder function.
Changing the two variables $x, E$ into $z,\nu$ as
\begin{align}
    z \equiv e^{+i\pi/4}x, \quad
    E \equiv +i\left( \nu+\frac{1}{2} \right),
\end{align}
the general form of the solution is written as
\begin{align}
    \psi = A\cdot D_{\nu}(z) + B\cdot D_{-\nu-1}(iz),
\end{align}
where $A,B$ are some constants.

\begin{figure}[t]
    \centering
    \includegraphics[width=80mm]{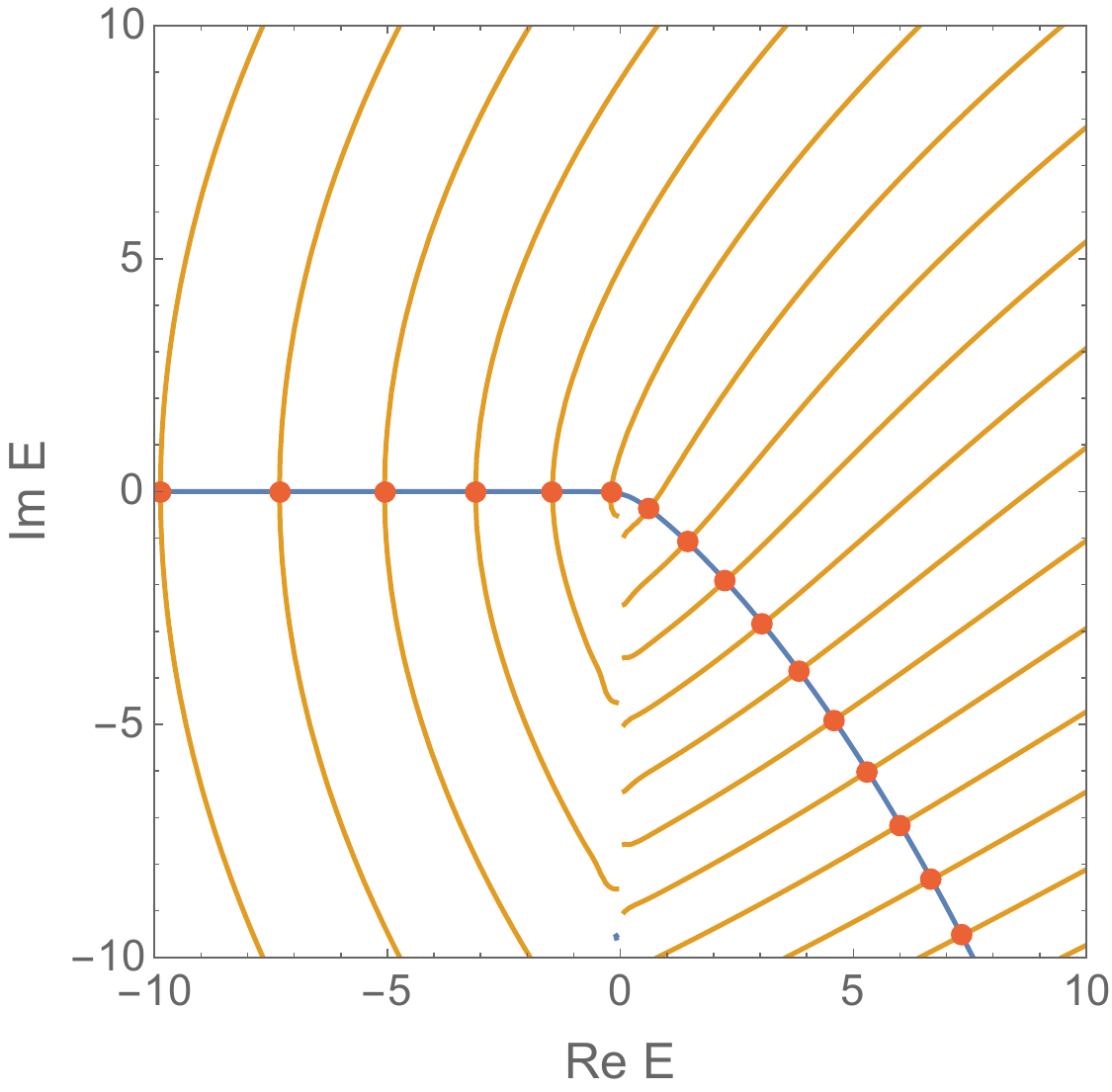}
    \caption{
	The spectrum of the inverted harmonic oscillator
        $\widetilde{V}(x)=-\frac{x^2}{4}$ (indicated as red blobs).
        We have imposed the Dirichlet boundary condition at $x=L=10$,
        and have assumed that there is only the out-going mode in the region $x\rightarrow-\infty$.
        The blue curve represents $\Im\vartheta_+=0$,
        while the orange curve represents $\Re\vartheta_+\in 2\pi\mathbf{Z}$.
        The intersection points of the two curves represent the quantized energy spectrum.
    }
    \label{fig:IHO_ESpec_L10}
\end{figure}

The parabolic cylinder function $D_{\nu}(z)$ has an asymptotic form
\begin{align}
    \label{eq:para_cylinder_asymp}
    D_{\nu}(z)
    \sim \left\{
    \begin{array}{ll}
        e^{-z^2/4} z^{\nu}
        -\dfrac{\sqrt{2\pi}}{\Gamma(-\nu)} e^{\pm i\pi\nu}
        e^{z^2/4} z^{-\nu-1} &
        \;\;\; \pi/4 < \pm\arg z < 5\pi/4 \\
        e^{-z^2/4} z^{\nu} &
        \;-3\pi/4 < \arg z < 3\pi/4
    \end{array}
    \right.
\end{align}
For our later convenience, we introduce a function
\begin{align}
    \vartheta_{\pm}(\nu;z)
    &\equiv  \frac{-iz^2}{2} + i(2\nu+1)\ln z
    \pm\pi\nu -i\ln\frac{\sqrt{2\pi}}{\Gamma(-\nu)}
\end{align}
so that the location of zeros of the wave function is easily spotted,
\begin{align}
    D_{\nu}(z)
    &\sim e^{-z^2/4}z^{\nu}\: (1-e^{i\vartheta_{\pm}(\nu;z)}), \quad
    \pi/4 < \pm\arg z < 5\pi/4.
\end{align}
By using the form of the asymptotic expansion,
we find that the parabolic cylinder function $D_{\nu}(z)$
has an in-going mode in the region $x\rightarrow+\infty$
and has in/out-going modes in the region $x\rightarrow-\infty$,
whereas $D_{-\nu-1}(iz)$
has in/out-going modes in the region $x\rightarrow+\infty$
and has an out-going mode in the region $x\rightarrow-\infty$.
Thus the appropriate choice of the solution is
\begin{align}
    \psi = B\cdot D_{-\nu-1}(iz).
\end{align}
Near the boundary $x \sim L \gg 1$,
the wave function takes the following form,
\begin{align}
    \psi
    \sim e^{-(iz)^2/4}(iz)^{-\nu-1}
    \left( 1-e^{i\vartheta_+(-\nu-1;iz)} \right).
\end{align}
Thus the Dirichlet boundary condition $0=\psi(L)$ yields a quantization condition,
\begin{align}
    \vartheta_+(-\nu-1;iz)
    \in 2\pi \mathbf{Z}.
\end{align}
That is, more explicitly,
\begin{align}
     \frac{\pi}{2} +\frac{L^2}{2}
    +\left( 2\ln L + \frac{\pi i}{2} \right)E
    +i\ln\frac{\sqrt{2\pi}}{\Gamma(1/2-iE)}
    \in 2\pi \mathbf{Z}.
    \label{eq:theta_2piZ}
\end{align}
The spectrum in the complex $E$-plane is plotted in Fig.~\ref{fig:IHO_ESpec_L10}.
The blue/orange curves are for 
$\Im\vartheta_+=0, \Re\vartheta_+\in 2\pi\mathbf{Z}$, respectively.
The intersection points of the two curves are the solutions of the quantization condition.

The solutions in the region $\Re E<0$ have almost vanishing imaginary part, and correspond to extremely stable quasinormal modes localized in the region between the photon sphere potential and the AdS boundary.
The existence of these modes were already pointed out by \cite{Festuccia:2008zx}, and 
our result is consistent with that.

The spectrum in the region $\Re E>0$ and $\Im E<0$ corresponds to unstable quasinormal modes above the top of the potential hill of the photon sphere.
The whole spectrum is located on the following peculiar curve in the complex $E$-plane,
\begin{align}
    \label{eq:IHO_Spec_Curve}
    0 = \Im\left[
        \left( 2\ln L + \frac{\pi i}{2} \right)E
        +i\ln\frac{\sqrt{2\pi}}{\Gamma(1/2-iE)}
    \right].
\end{align}
This is a curve whose negative imaginary part of $E$ grows when the real part of $E$ grows, see Fig.~\ref{fig:IHO_ESpec_L10}.

Let us consider the large angular momentum limit $l\to\infty$. 
As seen in \eqref{defL}, it corresponds to the large $L$ limit in the original quantum mechanical model.
When the boundary is sent to the spatial infinity, $L\rightarrow+\infty$, 
the slope of the curve \eqref{eq:IHO_Spec_Curve} in the $E$-plane decreases and the curve approaches the positive real axis. 
In fact, expanding the gamma function $\Gamma(1/2-iE)$ around $E=0$,
we find
\begin{align}
    \label{eq:IHO_spec_largeL}
    E\sim
    \dfrac{\pi}{\ln L}
    \left\{
        n -\left(
            \frac{L^2}{4\pi} - \left\lfloor\frac{L^2}{4\pi}\right\rfloor
        \right) -\frac{1}{4}
        -\frac{i\ln 2}{4\pi}
    \right\}, \quad
    n= 1,2,\dots \ll \ln L.
\end{align}
The quantum number $n$ counts the QNM, and the imaginary part of the spectrum $E$ becomes a constant negative value independent of $n$.

Before closing this subsection,
it will be worthwhile to compare this spectrum to the one with different boundary conditions.
Suppose that we instead allow only out-going modes on both sides of the potential hill, $x\rightarrow\pm\infty$.
The spectrum is obtained by killing the in-going mode of $\psi = B\cdot D_{-\nu-1}(iz)$ in the region $x\rightarrow+\infty$ by setting
\begin{align}
    -\frac{\sqrt{2\pi}}{\Gamma(1+\nu)}e^{+i\pi(-1-\nu)} = 0.
\end{align}
The gamma function has poles at non-positive integers, thus,
\begin{align}
    \label{eq:IHO_spec_wo_bdy}
    E
    = -i\left(n+\frac{1}{2}\right), \quad
    n=0,1,2,\dots
\end{align}
The apparent mismatch between \eqref{eq:IHO_spec_wo_bdy} and \eqref{eq:IHO_spec_largeL} in the limit $L\rightarrow\infty$ is not a contradiction.
In our case of \eqref{eq:IHO_spec_largeL},
for a decaying solution $\Im E<0$,
the out-going mode of $D_{-\nu-1}(iz)$ grows in power as $x$ approaches the boundary,
whereas the in-going mode decreases in power.
Thus we needed to add an in-going mode with a large amplitude coefficient to realize the Dirichlet boundary condition.
This is different from just having the out-going mode. Therefore, the existence of the hard wall, even placed at the spatial infinity, discontinuously changes the spectrum.

\subsection{P\"{o}schl-Teller model}

In this subsection, we demonstrate that the curve \eqref{eq:IHO_Spec_Curve} is almost insensitive to the shape of the potential tail in a solvable model.
We replace the potential of the quantum mechanical model \eqref{QM} with that of the P\"{o}schl-Teller type,
\begin{align}
    \widetilde{V}(x)
    \rightarrow V_0\left(\frac{1}{\cosh^2\alpha x}-1\right).
\end{align}
The top of the hill in this potential is approximated by the inverted harmonic potential, while the potential has a potential tail which resembles the actual shape of the potential near the AdS boundary, \eqref{Vuni}.
For our later convenience, we introduce a variable
\begin{align}
    \omega^2 = V_0 + E,
\end{align}
and consider the Schr\"{o}dinger equation
with the P\"{o}schl-Teller potential\footnote{See App.~\ref{sec:B-2} for more detailed discussion on the spectrum.}
\begin{align}
	\left[
	\dv[2]{x} + \omega^2
        -\frac{V_0}{\cosh^2\alpha x}
	\right]
	\psi(x)
	= 0,
\end{align}
where
\begin{align}
    &V_0 \equiv  \alpha^2 \lambda(1-\lambda), \\
    &\alpha > 0, \quad
    \lambda \equiv  \frac{1}{2} + i\nu, \quad \nu > 0.
\end{align}
In a manner similar to the inverted harmonic oscillator,
we shall solve this Schr\"{o}dinger equation
under the Dirichlet boundary condition,
\begin{align}
    \psi(L) = 0, \quad
    L \gg 1.
\end{align}
Also we assume that
the wave function $\psi(x)$ has only the out-going mode
in the region $x\rightarrow-\infty$, under the condition that the wave function decays with respect to time, $\Im E=\Im\omega^2<0$.

\begin{figure}[t]
    \centering
    \includegraphics[width=75mm]{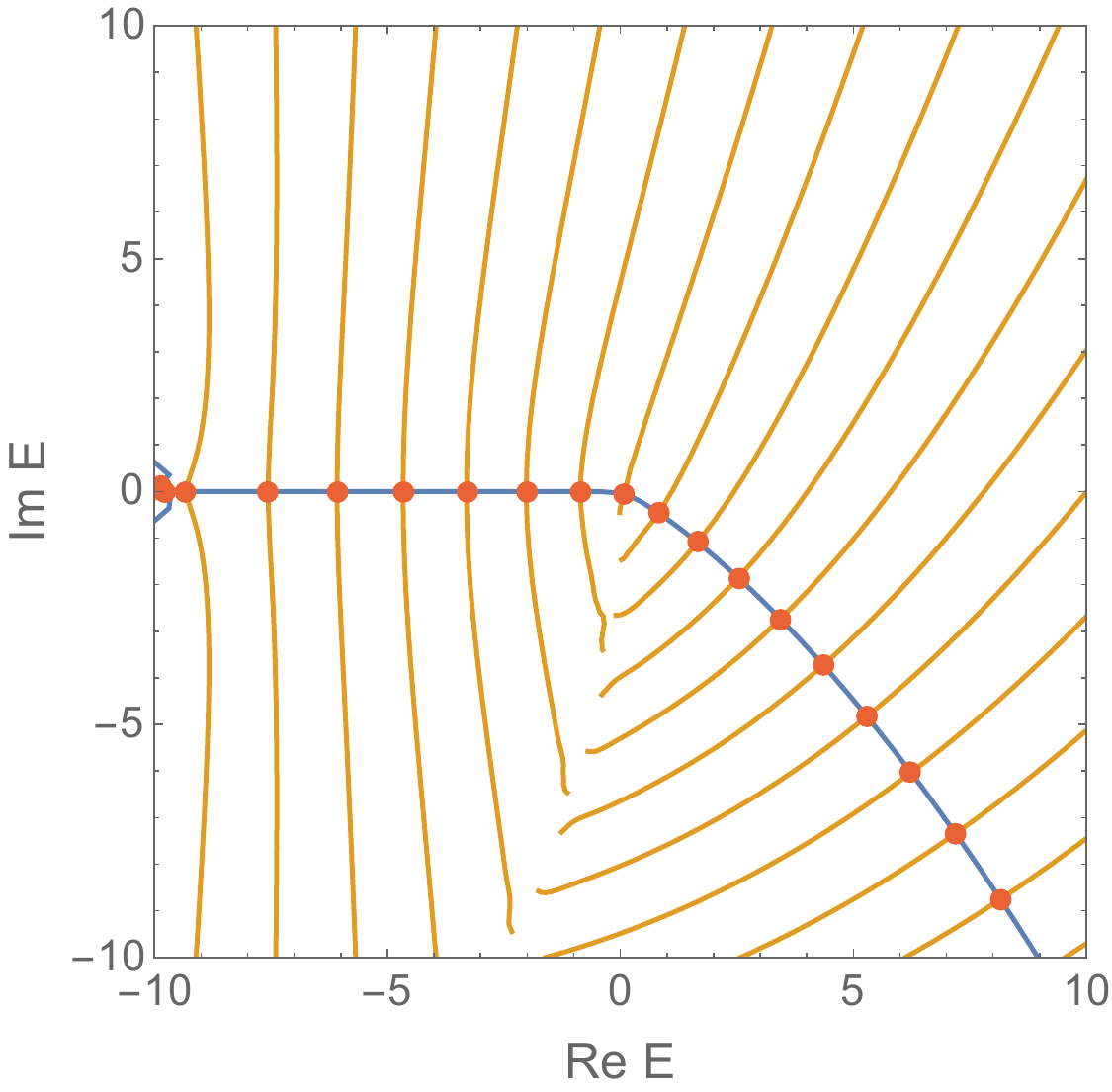}
    \includegraphics[width=75mm]{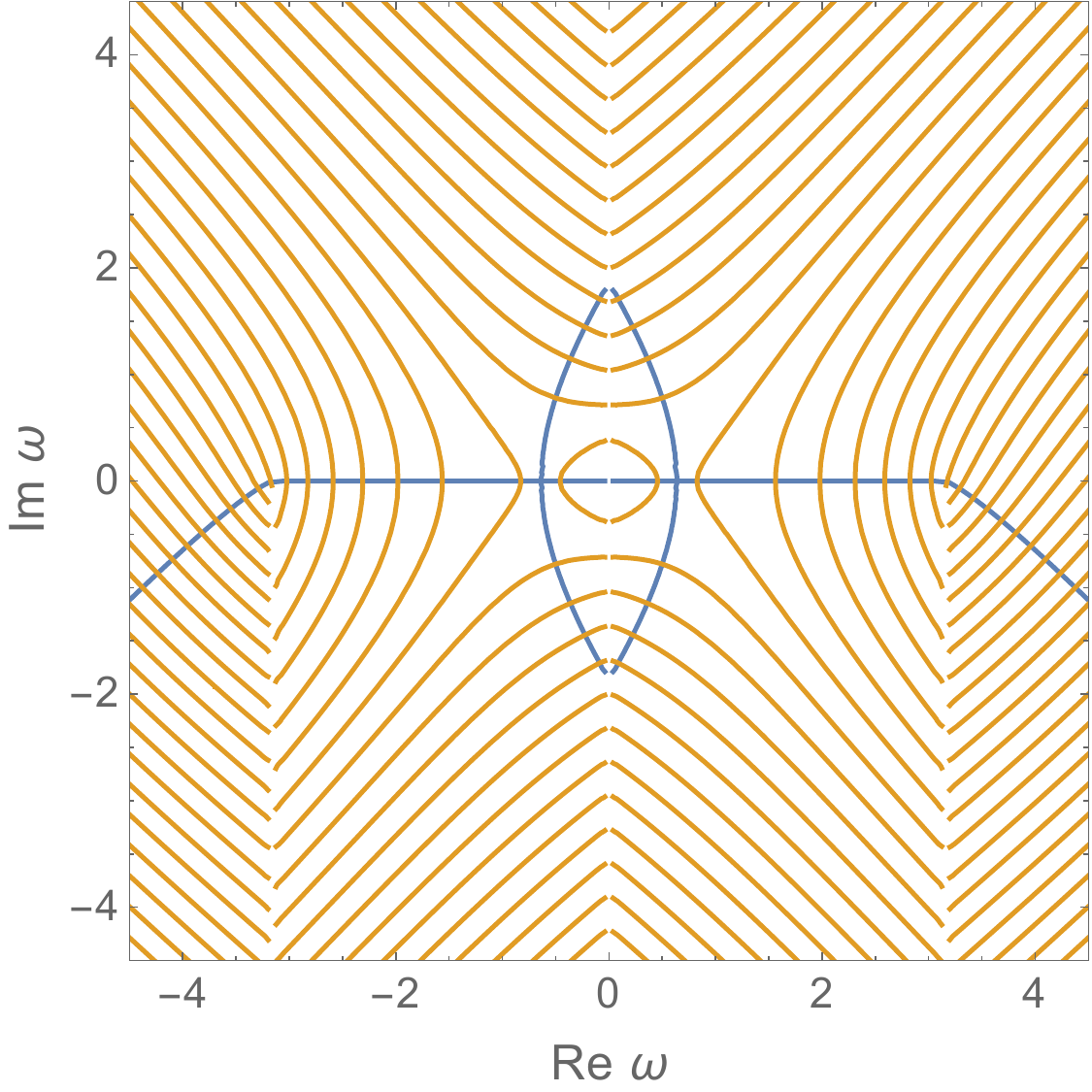}
    \caption{
	The spectrum of the P\"{o}schl-Teller type potential $\widetilde{V}(x)=V_0\left(\frac{1}{\cosh^2\alpha x}-1\right)$ with $V_0=10$ (indicated as red blobs).
        We have imposed the Dirichlet boundary condition at $x=L=10$,
        and have assumed that there is only the out-going mode in the region $x\rightarrow-\infty$.
        Other parameters are tuned as
        $\alpha=1/\sqrt{4V_0},\: \nu=\sqrt{4V_0^2-1/4}$
        so that we keep $\widetilde{V}(x)\simeq-\frac{x^2}{4}$ around the top of the potential hill, to be compared with Fig.~\ref{fig:IHO_ESpec_L10}.
        The blue curve represents $\Im\vartheta=0$,
        while the orange curve represents $\Re\vartheta\in 2\pi\mathbf{Z}$.
        Their intersection points represent the quantized energy spectrum. The left (right) panel is the spectrum in the $E$-plane ($\omega$-plane).
    }
    \label{fig:PT_ESpec_L10_V10}
\end{figure}

The Schr\"{o}dinger equation
has two independent solutions
(see Eq.(9) in \cite{Cevik:2016mnr}):
\begin{align}
    f_+(x)
    &= e^{i\omega x}
    \F{\lambda}{1-\lambda}{1+\frac{i\omega}{\alpha}}
    {\frac{e^{\alpha x}}{ e^{\alpha x}+e^{-\alpha x} }}, \\
    f_-(x)
    &= (e^{\alpha x}+e^{-\alpha x})^{\frac{i\omega}{\alpha}}
    \F{\lambda-\frac{i\omega}{\alpha}}{1-\lambda-\frac{i\omega}{\alpha}}
    {1-\frac{i\omega}{\alpha}}
    {\frac{e^{\alpha x}}{ e^{\alpha x}+e^{-\alpha x} }}.
\end{align}
Here $F$ is the hypergeometric function.
Transforming variables, we find
\begin{align}
	f_+(x)
        &= e^{i\omega x}
	\frac{\Gamma(1+i\omega/\alpha)\Gamma(i\omega/\alpha)}
	{\Gamma(\lambda+i\omega/\alpha)\Gamma(1-\lambda+i\omega/\alpha)}
	\F{\lambda}{1-\lambda}{1-i\omega/\alpha}
	{\frac{1}{e^{2\alpha x}+1}} \notag\\
	&
	+(e^{\alpha x}+e^{-\alpha x})^{-i\omega/\alpha}
	\frac{\Gamma(1+i\omega/\alpha)\Gamma(-i\omega/\alpha)}
	{\Gamma(\lambda)\Gamma(1-\lambda)}
	\F{\lambda+i\omega/\alpha}{1-\lambda+i\omega/\alpha}
	{1+i\omega/\alpha}
	{\frac{1}{e^{2\alpha x}+1}} \notag\\
        &\sim \left\{
        \begin{array}{ll}
            e^{i\omega x}
            & \quad (\Im\omega>0, x\rightarrow -\infty) \\
            e^{i\omega x}
    	\dfrac{\Gamma(1\!+\!i\omega/\alpha)\Gamma(i\omega/\alpha)}
            {\Gamma(\lambda\!+\!i\omega/\alpha)\Gamma(1\!-\!\lambda\!+\!i\omega/\alpha)}& \\
            \hspace{20mm}+ e^{-i\omega x}
            \dfrac{\Gamma(1\!+\!i\omega/\alpha)\Gamma(-i\omega/\alpha)}
            {\Gamma(\lambda)\Gamma(1\!-\!\lambda)}
            & \quad (\Im\omega>0, x\rightarrow +\infty)
        \end{array}
        \right. \\
	f_-(x)
        &= (e^{\alpha x}+e^{-\alpha x})^{i\omega/\alpha}
	\frac{\Gamma(1-i\omega/\alpha)\Gamma(i\omega/\alpha)}
	{\Gamma(\lambda)\Gamma(1-\lambda)}
	\F{\lambda-i\omega/\alpha}{1-\lambda-i\omega/\alpha}
	{1-i\omega/\alpha}
	{\frac{1}{e^{2\alpha x}+1}} \notag\\
	&\qquad\quad
	+e^{-i\omega x}
	\frac{\Gamma(1-i\omega/\alpha)\Gamma(-i\omega/\alpha)}
	{\Gamma(\lambda-i\omega/\alpha)\Gamma(1-\lambda-i\omega/\alpha)}
	\F{\lambda}{1-\lambda}
	{1+i\omega/\alpha}
	{\frac{1}{e^{2\alpha x}+1}} \notag\\
        &\sim \left\{
	\begin{array}{ll}
		e^{-i\omega x}
            &  \quad (\Im\omega<0, x \rightarrow -\infty) \\
		e^{i\omega x}
		\dfrac{\Gamma(1\!-\!i\omega/\alpha)\Gamma(i\omega/\alpha)}
		{\Gamma(\lambda)\Gamma(1\!-\!\lambda)} & \\
		\hspace{20mm}+ e^{-i\omega x}
		\dfrac{\Gamma(1\!-\!i\omega/\alpha)\Gamma(-i\omega/\alpha)}
		{\Gamma(\lambda\!-\!i\omega/\alpha)\Gamma(1\!-\!\lambda\!-\!i\omega/\alpha)}
		& \quad (\Im\omega<0, x \rightarrow +\infty)
	\end{array}
	\right.
\end{align}
Thus the appropriate choice of the solution is
\begin{align}
    \psi \propto f_-(x).
\end{align}
Near the boundary $x\sim L \gg 1$,
the wave function takes the following form
\begin{align}
    \psi
    \sim e^{i\omega x}
	\dfrac{\Gamma(1-i\omega/\alpha)\Gamma(i\omega/\alpha)}
	{\Gamma(\lambda)\Gamma(1-\lambda)}\:
	\left( 1 - e^{i\vartheta(\omega;x)} \right),
\end{align}
where
\begin{align}
	\vartheta(\omega; x)
	\equiv  \pi - 2\omega x
	-i \ln \frac{\Gamma(-i\omega/\alpha)}{\Gamma(i\omega/\alpha)}
	\frac{\Gamma(\lambda)\Gamma(1-\lambda)}
	{\Gamma(\lambda-i\omega/\alpha)\Gamma(1-\lambda-i\omega/\alpha)}.
 \label{eq:thetaPTdef}
\end{align}
The Dirichlet boundary condition $0=\psi(L)$
yields the following quantization condition,
\begin{align}
     \vartheta(\omega,L) \in 2\pi\mathbf{Z}.
     \label{eq:theta_2pi_Z_PT}
\end{align}
Its spectrum is plotted in the left panel of Fig.~\ref{fig:PT_ESpec_L10_V10}.
The blue/orange curves are for
$\Im\vartheta=0,\; \Re\vartheta\in 2\pi\mathbf{Z}$,
respectively.
Their intersection points are the solutions of the quantization condition.
The parameters in the potential are chosen so that $\widetilde{V}(x)\simeq -\frac{x^2}{4}$ around the top of the potential hill,
to be compared with the result of the inverted harmonic oscillator, Fig.~\ref{fig:IHO_ESpec_L10}.

As in the case of the inverted harmonic oscillator,
the spectrum in the region $\Re E<0$ has almost vanishing imaginary part, and corresponds to extremely stable quasinormal modes living in the spatial region between the photon sphere potential and the AdS boundary.
Other spectrum in the region $\Re E>0$ and $\Im E<0$ corresponds to unstable modes above the top of the potential hill of the photon sphere.
Therefore, we conclude that the spectral pattern of this model shares the same characteristic with that of the inverted harmonic potential.

In the right panel of Fig.~\ref{fig:PT_ESpec_L10_V10}, we plot the spectrum in the $\omega$-plane, for the readers' reference.
In the $\omega$-plane, spectra appear on both sides of positive/negative real axis since there are two branches $\omega=\pm\sqrt{E+V_0}$.
One of the blue horizontal curves in the $\omega$-plane is mapped to the blue curve on $E$-plane.\footnote{
As a remark aside, one observes other series of complex solutions in the $\omega$-plane: the intersection points of two blue vertical curves and series of orange curves around the origin, see the central part of the right panel of Fig.~\ref{fig:PT_ESpec_L10_V10}.
Such a series of solutions is mapped to the region $\Re E<-V_0=-10$,
which is below the bottom of the potential defined by the potential tail.
Thus it is natural to regard it as an unphysical series of solutions and 
we ignored it.
It would be interesting if this unphysical series of solutions allow some holographic interpretation, in quantum path-integral.
}

We can show that the equation for the blue curve $\Im\vartheta=0$ approaches that of the inverted harmonic oscillator when the height of the potential $V_0$ is sent to the infinity.
Let us fix the parameters as
\begin{align}
    \alpha = \frac{1}{\sqrt{4V_0}}, \quad
    \nu = \sqrt{4V_0^2-\frac{1}{4}}
\end{align}
so that $\widetilde{V}(x)\simeq-\frac{x^2}{4}$ near the top of the potential hill,
then send the potential height to a large value.
Some straightforward calculations of \eqref{eq:thetaPTdef} with the Stirling's formula show that
\begin{align}
    \vartheta(\omega;x)
    &\sim
    \text{(Real const.)}
    -\left( 2\ln L_{\text{eff.}}+\frac{\pi i}{2} \right)E
    -i\ln\frac{\sqrt{2\pi}}{\Gamma(1/2-iE)},
\end{align}
where 
 $ 2 \ln L_{\text{eff.}} \equiv \frac{L}{\sqrt{V_0}}+\ln V_0$.
Thus the curve on which quasinormal modes are located is
\begin{align}
    \label{eq:PT_Spec_Curve}
    0 = \Im\left[
        \left( 2\ln L_{\text{eff.}} + \frac{\pi i}{2} \right)E
        +i\ln\frac{\sqrt{2\pi}}{\Gamma(1/2-iE)}
    \right],
\end{align}
which is exactly the same form as \eqref{eq:IHO_Spec_Curve}.

Our results in the inverted harmonic oscillator and in the P\"{o}schl-Teller type potential imply that quasinormal modes near the photon sphere appear with a peculiar spectral pattern, in the large angular momentum sector. The pattern is shown in Fig.~\ref{fig:IHO_ESpec_L10} and Fig.~\ref{fig:PT_ESpec_L10_V10}, and it is on a curve of the form given by \eqref{eq:IHO_Spec_Curve}.

\section{Summary and discussion}
\label{sec:5}

In this paper, we calculated the spectrum $\Omega$ of the scalar quasinormal mode (QNM) associated with the photon sphere in the AdS Schwarzschild black hole.
In the large $l$ limit where $l$ is the angular momentum of the QNM, we found that $\Omega$ is related to the energy spectrum $E$ of the quantum mechanics (QM) \eqref{QM}, as given in \eqref{omegaE}. At high temperature, the QM potential was analytically calculated as \eqref{Vuni}. The QM is not solvable analytically, so we instead analytically solved the inverted harmonic oscillator with a hard wall as \eqref{eq:theta_2piZ}, to obtain the feature of the QM spectrum $E$.
The resultant $E$ is a discrete series along a curve on the complex $E$-plane. As we see in Fig.~\ref{fig:IHO_ESpec_L10}, $\Im E$ (which needs to be negative as it is for decaying modes) grows in its magnitude as Re$E$ grows, while $\Im E$ almost vanishes for negative $\Re E$.  
Quite a similar pattern of the spectrum is universally obtained in the P\"oschl-Teller model with a hard wall (see Fig.~\ref{fig:PT_ESpec_L10_V10}), which suggests that the QM has the same spectral feature, which is translated to the feature of the QNMs through the relation \eqref{omegaE}.
This spectrum, by the AdS/CFT dictionary, should be equal to the spectrum of a thermal holographic CFT on a sphere whose radius is $l_0$, at the large angular momentum $l$.

An interesting feature of the obtained QNM spectrum, \eqref{omegaE} with \eqref{eq:theta_2piZ} or \eqref{eq:theta_2pi_Z_PT}, is the fact that the spectrum whose real part is larger than the photon sphere energy always accompanies a non-vanishing imaginary part which is not directly related to the temperature.
Normally the imaginary part of QNMs is due to the boundary condition at the black hole horizon, thus is a direct consequence of the temperature. In the present case, the imaginary part is dictated by the photon sphere, which is determined mainly by the angular momentum.
The precise spectrum \eqref{omegaE} is a prediction to the thermal holographic CFT, and in general holographic QFTs we expect that such a ``photon-sphere subsector" exists in their spectrum.

Here note also that the spectrum whose real part is smaller than the photon sphere energy ($\Re E<0$) has almost vanishing imaginary part,
meaning that they are extremely stable. These modes are confined in the radial region bounded by the photon sphere and the AdS boundary. The existence of these modes were already pointed out in \cite{Festuccia:2008zx}, and our result is consistent with it.

Let us discuss a dual CFT interpretation of the QNMs which we analyzed in this paper. They should correspond to an operator of the form tr$[FDDD\cdots DDF]$ where $D=D_\theta$ is the covariant derivative along the equator of the sphere. This type of operator carries a large angular momentum on the sphere.
In our analysis, the spectrum consists of (i) those at $\Re E>0$ with nonzero and growing imaginary part, and (ii) those at $\Re E<0$ with almost vanishing imaginary part.

The latter (ii) corresponds to the stable states \cite{Festuccia:2008zx}, which are localized near the AdS boundary and spinning along the boundary sphere.
We see that their energy $\Omega$ is given as $l/l_0$ where $l$ is the angular momentum (and thus the number of $D_\theta$'s in the operator) and $l_0$ is the radius of the sphere. This is reminiscent of the pp-wave limit and the BMN operators \cite{Berenstein:2002jq}, which are stable and dictated by a long spin chain. The stability of these states suggests a bound between the angular momentum and the energy, which reminds us of the Kerr bound for black hole masses. The bound could be explained by some symmetry; in fact, as is seen in the effective potential \eqref{Vasymp}, the system is approximated by a harmonic oscillator around the AdS boundary $r_*=0$, with the Dirichlet boundary condition at $r_*=0$. This means that the spectrum of the QNMs localized near the AdS boundary is dictated by an $SL(2;\mathbf{R})$ symmetry (which should also act on the operator space in the CFT). Further study on these states would be interesting.

On the other hand, what we studied in detail in this paper is the sector (i) whose real part of the energy $\Omega$ is around the photon sphere, $\Re \Omega\sim l\sqrt{1/l_0^2 + 1/r_0^2}$. This value of the energy is at the top of the potential hill. The QNM develops the imaginary part. The corresponding operator in the CFT side is of the same form, tr$[FDDD\cdots DDF]$, but with very many impurity operators inserted. When the number of the impurities exceeds the photon sphere bound $\sim l \cdot l_0^2/r_0^2$ (which amounts to the energy difference between the top of the potential hill and the bottom at the AdS boundary), the states becomes suddenly unstable and decays.
In other words, the state with the large angular momentum of the form tr$[FDDD\cdots DDF]$ could be quite stable against large addition of impurities whose number can go up to $\sim l \cdot l_0^2/r_0^2$. This universal behavior of the CFT states, expected through the AdS/CFT duality, may need more intuitive explanation in the CFT language.

In the following, we present several discussions on how our work may be related to some other aspects in AdS/CFT and holography.

\begin{itemize}
\item 
First, let us discuss the relation between our spectra and those found in the asymptotically flat case \cite{Hadar:2022xag}. As we have emphasized in Sec.~\ref{sec:4}, our AdS case has a hard wall at the AdS boundary, while the asymptotically flat case does not have the wall. This difference was crucial in determining the spectrum. Now, suppose that we add a flat spacetime joined to the AdS boundary, as was done in \cite{Almheiri:2019qdq}. Then the wall could ``disappear" and the QNM in the AdS bulk can escape to the joined flat spacetime. Since the AdS plus the joined flat space may allow a holographic interpretation as a CFT joined with a higher dimensional CFT, the spectrum found in \cite{Hadar:2022xag}, which is dictated by the $SL(2;\mathbf{R})$ symmetry, may be realized in the AdS/CFT setup in this manner. It would be interesting to find the AdS/CFT interpretation of the $SL(2;\mathbf{R})$ subsector observed in \cite{Hadar:2022xag}, in view of the issues concerning evaporation of black holes and the Page curve.

\item
Einstein rings can be obtained in holographic CFT \cite{Hashimoto:2018okj,Hashimoto:2019jmw} by using an imaging transform on a CFT one-point function.\footnote{The imaging in holography and string theory has been further developed for various purposes, see Refs.~\cite{Kaku:2021xqp,Liu:2022cev,Kaku:2022hcg,Hashimoto:2022ugt,Mandal:2022oma,Caron-Huot:2022lff,Hashimoto:2022aso,Zeng:2023zlf,Kinoshita:2023hgc}.}
The CFT is with a point-like source, thus the holographic imaging of the Einstein rings is a direct consequence of two-point functions on the holographic CFT. 
Since generally the QNM spectrum near the photon sphere should characterize Einstein rings, it would be interesting to find a relation between the image of the holographic Einstein rings and the photon-sphere subsector of the CFT spectrum.

\item
In the context of the AdS/CFT, Lyapunov exponents have been studied in connection to the chaos bound saturation \cite{Maldacena:2015waa}, as it is related to the black hole surface gravity \cite{Shenker:2013pqa,Engelsoy:2016xyb,Hashimoto:2016dfz}. The QNMs associated with the photon sphere naturally defines the Lyapunov exponent there \cite{Cardoso:2008bp}, thus its relevance to the chaos bound would be interesting if exists. Namely, the chaos bound could work as a bound for the QNM spectra. This research direction would contribute to some general discussion on the black hole instability, such as \cite{Bizon:2011gg}. 

\end{itemize}

Further interplay between boundary CFTs and photon spheres will reveal the mystery of the AdS/CFT correspondence.

\subsection*{Acknowledgment}

We would like to thank K.~Yoshida for valuable discussions.
The work of K.~H.\ was supported in part by JSPS KAKENHI Grant No.~JP22H01217, JP22H05111 and JP22H05115.
The work of K.~S.\ was supported in part by Grant-in-Aid for JSPS Fellows No.~23KJ1310.
The work of T.~Y.\ was supported in part by JSPS KAKENHI Grant No.~JP22H05115 and JP22KJ1896.

\appendix


\newcommand{\beq}{\begin{eqnarray}}
\newcommand{\eeq}{\end{eqnarray}}
\newcommand{\del}{\partial}

\renewcommand{\ket}[1]{| {#1} \rangle}
\renewcommand{\bra}[1]{ \langle {#1} | }
\newcommand{\geom}[4]{F \left(\begin{array}{c} {#1}, {#2} \\ {#3}  \end{array} ; {#4} \right)}

\section{Representation of \texorpdfstring{$SL(2;{\bf R})$}{TEXT}}
\label{sec:A}

We review several properties of  $SL(2;{\bf R})$ and 
discuss the representation of its Lie algebra. 

\subsection{Relation between \texorpdfstring{$SO(2,1), SU(1,1)$}{TEXT} and \texorpdfstring{$SL(2;{\bf R})$}{TEXT}}

There are three kinds of noncompact Lie algebras 
$so(2,1) \cong su(1,1) \cong sl(2,{\bf R})$ that are isomorphic one another as described in the Fig.~\ref{so-su-sl}. They play important roles in physics.

The noncompact group $SL(2;{\bf R})$ is the linear group consisting of 
matrices $g=\left(\begin{array}{cc} \alpha & \beta \\ 
\gamma & \delta \\ \end{array}\right)$
( $\alpha ,\beta ,\gamma ,\delta \in{\bf R}$ ) with a relation $\alpha \delta -\beta \gamma =1$. The $SU(1,1)$ is the complex linear group 
generated by $g'=\left(\begin{array}{cc} a & b \\ 
\bar{b} & \bar{a} \\ \end{array}\right)$
$(a,b\in {\bf C})$ with a constraint $|a|^2-|b|^2=1$.

This $SU(1,1)$ is isomorphic to $SL(2;{\bf R})$ 
under a similarity transformation 
$g=t^{-1}\cdot g' \cdot t $ for
$g \in SU(1,1) $ and $g' \in SL(2;{\bf R})$ with 
$t\equiv \frac{1}{\sqrt{2}}
\left(
\begin{array}{cc}
 1 & i \\
i & 1 \\
\end{array}
\right) $. 
That leads to relations of entries of matrices as
$a=\frac{\alpha +\delta }{2}+i \frac{\beta -\gamma}{2}$, 
$b=\frac{\beta +\gamma }{2}+i \frac{\alpha -\delta}{2}$. 
As concrete examples, we can consider 
1-parameter subgroups of $SL(2;{\bf R})$ and their correspondents as $SU(1,1)$ elements:
\beq
&& \mbox{elliptic}\,;\,
\left(
\begin{array}{cc}
\cos \frac{t}{2} & -\sin \frac{t}{2} \\
\sin \frac{t}{2} & \cos \frac{t}{2} \\
\end{array}
\right) \leftrightarrow 
\left(
\begin{array}{cc}
e^{-i \frac{t}{2}} & 0 \\
0 &  e^{+i\frac{t}{2}} \\
\end{array}
\right)\,,\\
&& \mbox{hyperbolic}\,;\,
\left(
\begin{array}{cc}
e^{-t/2} & 0 \\
0 & e^{t/2} \\
\end{array}
\right)
\leftrightarrow \left(
\begin{array}{cc}
\cosh \frac{t}{2} & -i \sinh \frac{t}{2} \\
i \sinh \frac{t}{2} & \cosh \frac{t}{2} \\
\end{array}
\right)\,,\\
&&
\mbox{parabolic}\,;\,
\left(
\begin{array}{cc}
1 &  0 \\
t & 1 \\
\end{array}
\right)
\leftrightarrow
\left(
\begin{array}{cc}
1-i \frac{t}{2} &  \frac{t}{2} \\
 \frac{t}{2} & 1+i \frac{t}{2} \\
\end{array}
\right)\,.
\eeq

\begin{figure}

\beq
\begin{array}{ccccc}
SO(3,1) & \leftrightarrow & SL(2;{\bf C}) & & \\
\downarrow & & \downarrow & & \\
SO(2,1) & \leftrightarrow & SU(1,1) & \leftrightarrow & SL(2;{\bf R}) \\
\end{array}
\eeq

\centering

\caption{Non compact groups are related one another. $SO(2,1)$ is constructed 
by a reduction from $SO(3,1)$. $SU(1,1)$ is obtained from $ SL(2;{\bf C}) $ by a suitable restriction. $ SU(1,1) $ is also related to $SL(2;{\bf R})$ by a similarity transformation.}

\label{so-su-sl}

\end{figure}

The $SU(1,1)$ is a covering space of $SO(2,1)$, 
and the latter leaves the bilinear form $x_0^2-x_1^2-x_2^2$ invariant and 
is a subgroup of a Lorentz group $SO(3,1)$. 
The center of $SU(1,1)$ is ${\bf Z}_2$, and 
$SU(1,1)/{\bf Z}_2$ is locally isomorphic to 
$SO(2,1)$. Indeed
$SU(1,1)$ is a simply connected group and is 
the covering group of $SO(2,1)$.

$SO(3,1)$ has a covering space $SL(2,{\bf C})$
and we can write down relations among components 
of $SU(1,1)$ and $SL(2,{\bf C})$ 
by using a reduction of $SO(3,1)$ to $SO(2,1)$.

The group $SO(3,1)$ acts on a set of coordinates 
$x^{\mu}=(x^0,x^1,x^2,x^3)$ and $x^{\mu}$ transforms into 
$x'{}^{\mu}=\Lambda^{\mu}{}_{\nu}x^{\nu}$ $(\Lambda \in SO(3,1))$. 
By introducing the Pauli matrices, one can represent
this transformation as a matrix form ${\bf x}=x^{\mu}\sigma_{\mu}$. 
The $SL(2,{\bf C})$ can act 
on this matrix as left/right  multiplications by an element 
$ \hat{g}\in SL(2,{\bf C})$. 

\beq
&& x^{\mu}=(x^0,x^1,x^2,x^3)\,,\,
 \sigma_{\mu}=(1,\sigma_1,\sigma_2, \sigma_3)\,,\\
&& {\bf x}:=x^{\mu}\sigma_{\mu}=
\left(
\begin{array}{cc}
x^0 +x^3& x^1-i x^2 \\
x^1+i x^2 & x^0-x^3 \\
\end{array}
\right)\,,\\
&&
\hat{g}=\exp \left(i\vec{\theta}\cdot \frac{\vec{\sigma}}{2}-\vec{\omega}\cdot 
\frac{\vec{\sigma}}{2}\right)\,,\,\,
\det \hat{g}=1\,,\\
&& 
x'{}^{\mu}=\Lambda^{\mu}{}_{\nu} x^{\nu}\,,\,\,
 {\bf x}'=\hat{g}\cdot {\bf x}\cdot \hat{g}^{\dagger}\,.
\eeq
We can construct $SO(2,1)$ group
by putting $x^3=0$ and restricting  $SO(3,1)$ into $SO(2,1)$.
Then the covering space $SL(2;{\bf C})$ 
is reduced to $SU(1,1)$.
But one has to impose a constraint $\hat{g} \cdot \sigma_3 
\cdot \hat{g}^{\dagger}=\sigma_3$ 
for $\hat{g} \in SL(2;{\bf C}) $ to remove 
mixing between  the $x^3$ part and the remaining 3-dimensinal parts. 
This condition leads to a restriction of components of 
$\hat{g} \in SL(2;{\bf C})$ and induces an element of $g' \in SU(1,1)$, 
which is parameterized by two complex parameters $a$, $b$ with 
a constraint $|a|^2-|b|^2=1$,
\beq
&& SL(2;{\bf C})\ni \hat{g}=
\left(
\begin{array}{cc}
 a & b \\
c & d \\
\end{array}
\right)\,,\,\,ad-bc=1\,,\,\, a,b,c,d \in {\bf C}\,,\\
&& SU(1,1)\ni g' =
\left(
\begin{array}{cc}
 a & b \\
\bar{b} & \bar{a} \\
\end{array}
\right)\,,\,\,|a|^2-|b|^2=1\,,\,\, a,b \in {\bf C}\,.
\eeq

Every element $g' \in S(1,1)$ satisfies a constraint
 $g' \cdot \sigma_3 \cdot g'^{\dagger}=\sigma_3$.
 It induces a natural reduction of $SL(2;{\bf C})$ to $SU(1,1)$ 
 associated with the reduction of $SO(3,1)$ to $SO(2,1)$. 
 Then we can write down the corresponding element 
 $\Lambda^{\mu}{}_{\nu}$ of 
 $SO(2,1)$ as
 \beq
 && SO(2,1)\ni \Lambda^{\mu}{}_{\nu}=
 \left(
 \begin{array}{ccc}
 |a|^2+|b|^2 & 2 \mbox{Re} (a \bar{b}) & 2 \mbox{Im} (a \bar{b}) \\
  2 \mbox{Re} (a b) &  \mbox{Re} (a^2+ b^2) &  
  \mbox{Im} (a^2-b^2) \\
   -2 \mbox{Im} (a b )  &  -\mbox{Im} (a^2 +b^2) &  
   \mbox{Re} (a^2 -b^2) \\     
 \end{array}
 \right)\,, 
 |a|^2-|b^2|=1\,.
 \eeq
 By recalling the relations between $SU(1,1)$ and $SL(2;{\bf R})$, 
 we can show an element $g\in SL(2;{\bf R})$ in this parameterization as
 \beq
 && SL(2;{\bf R})\ni g=
 \left(
 \begin{array}{cc}
 \mbox{Re}(a) +\mbox{Im}(b) &  \mbox{Re}(b) +\mbox{Im}(a) \\
  \mbox{Re}(b) -\mbox{Im}(a) &  \mbox{Re}(a) -\mbox{Im}(b) \\
 \end{array}
 \right)\,,\,\,
 |a|^2-|b|^2=1\,. 
 \eeq
 
 \subsection{Representation of \texorpdfstring{$su(1,1)$}{TEXT}}
 
 For the universal covering group associated 
 with locally isomorphic Lie groups, all its 
 irreducible representations are single-valued. 
Keeping this in our mind, we study representations of the Lie albegra 
$su(1,1)$.
 
The Lie algebra $su(1,1)$ is generated by 
a set of $J_a$'s $(a=1,2,3)$. 
The commutation relations are expressed in the form
\beq
&& [J_1,J_2]=-iJ_3\,,\,\,[J_2,J_3]=i J_1\,,\,\,
[J_3,J_1]=iJ_2\,,\\
&& J_a^{\dagger}=J_a\,. 
\eeq
By introducing $J_{\pm}$, we rewrite commutators of the generators as
\beq
&& J_{\pm}=J_1\pm i J_2\,,\,\, (J_{\pm})^{\dagger} = J_{\mp}\,,\\
&&
[J_+, J_-]=-2 J_3\,,\,\, [J_3,J_{\pm}]=\pm J_{\pm}\,,
\eeq
and a Casimir invariant is represented as a quadratic form
\beq
J^2=J_1^2+J_2^2-J_3^2=\frac{1}{2}(J_+J_- +J_-J_+ )-J_3^2  
= J_{\pm}J_{\mp}\pm J_3 -J_3^2\,.
\eeq
When we take a redefinition 
$ L_0=-J_3\,,\,\,L_{\pm 1}=J_{\pm}$, we 
can express the Lie algebra $sl(2,{\bf R})$
with generators $L_n$ $(n=0, \pm 1)$. Their commutation relations 
are written as   
$ [L_{+1},L_{-1}]=2 L_0$,
$[L_0,L_{\pm 1}]=\mp L_{\pm 1}$.

Now we shall consider representations of $su(1,1)$. 
The operators $J_3$ and $J^2$ are hermitian operartors and 
their eigenvalues are respectively real numbers $a\in {\bf R}$, 
$\Lambda \in {\bf R}$. 
The states of the Lie algebra are labelled by this set of 
eigenvlues $(\Lambda ,a)$,
\beq
&& J^2 \ket{\Lambda ,a}=\Lambda \ket{\Lambda ,a}\,,\,\,
J_3 \ket{\Lambda ,a}=a \ket{\Lambda ,a}\,.\,\,
\eeq
By using  $(J_{\mp})^{\dagger}J_{\mp}=J_{\pm}J_{\mp}
=J^2+J_3^2\mp J_3$, 
a useful condition is obtained as
\beq
&& J_{\pm }J_{\mp}\ket{\Lambda ,a} =[\Lambda +a(a\mp 1)]
\ket{\Lambda ,a}\,,\,\,\\
&& \Lambda +a(a\mp 1)\geq 0.
\eeq
Here we assumed the inner product is positive definite. It is 
the natural assumption in the unitary theories.

We can also show that 
the generator $J_+$ raises the eigenvalue $a$ by one 
(when it satisfies $\bra{\Lambda ,a'}J_+\ket{\Lambda ,a} \neq 0$) 
by considering the following equation
\beq
\bra{\Lambda ,a'}J_+\ket{\Lambda ,a} 
=\bra{\Lambda ,a'} [J_3 ,J_+] \ket{\Lambda ,a} 
=(a'-a)\bra{\Lambda ,a'}J_+\ket{\Lambda ,a}
\,.
\eeq
By using these two properties, we study 
unitary representations of $su(1,1)$.
The representation is labelled by $\Lambda$ of the Casimir invariant. 
We parameterise $\Lambda =-j(j+1)$ by $j\in {\bf C}$.
The Casimir is the hermitian operator and its eigenvalue 
must be a real number. The reality of $\Lambda$ 
leads to an important constraint
that $\Lambda$ is expressed in either of the following two series,
\beq
&& \Lambda \in {\bf R}\Rightarrow 
\left\{
\begin{array}{l}
 j = \alpha \,\,(\alpha \in {\bf R}) \\
 j=-\frac{1}{2}+i\beta \,\,(\beta \in {\bf R}) \,.\\
\end{array}
\right.
\eeq
Each series is respectively parameterised by a set of real numbers, $\alpha$ and $\beta$. 
One can show only these two are possible by a direct calculation with 
$j=\alpha +i\beta$ $(\alpha ,\beta \in {\bf R})$ as follows:
\beq
 \Lambda &&=-(\alpha +i\beta)[(\alpha +1)+i\beta]\nonumber \\
&&=
\{-\alpha (\alpha +1)+\beta^2 \}-i \beta (2\alpha +1)\in {\bf R}\, .
\eeq

\begin{figure}
\beq
&&
\begin{array}{|lcl|}\hline
\Lambda \leq 0 & & (j\geq 0)\\
 0< \Lambda \leq \frac{1}{4} & & (-\frac{1}{2}\leq j <0)\\
 0\leq \Lambda <\frac{1}{4} & & ( -1\leq j<-\frac{1}{2} ) \\
 \Lambda <0 & &  (j<-1) \\ \hline
\end{array}
\eeq

\centering

\caption{The values of $\Lambda$ and the range of $j$ are shown in the table. 
$\Lambda$ is negative or zero for the cases $j\geq 0$ or $j<-1$. 
On the other hand, $\Lambda$ is positive but is less than or equal to $\frac{1}{4}$ 
for the case $ -1\leq j <0$.}

\label{casimir}

\end{figure}

In the first case $j\in {\bf R}$, 
 the states are labelled by a set of real  numbers $(j,a)$.
$\Lambda$ is also found to be less than or equal to 
$\frac{1}{4}$ from the relation 
$\Lambda =\frac{1}{4}-(j+\frac{1}{2})^2\leq \frac{1}{4}$.
There are four regions of $j$ 
according to the value of $\Lambda$, as described in Fig.~\ref{casimir}.
However,
$\Lambda =-j(j+1)$ is invariant under 
a transformation $j\rightarrow -j-1$, so we can restrict 
$j$ to be negative, $j<0$. 
Then, in summary, the states are labelled by a set of real  numbers $(j,a)$ with $j<0$.

In the second case $j=-\frac{1}{2}+i\beta$ 
$(0\neq \beta \in {\bf R})$, 
a relation $\Lambda =\frac{1}{4} +\beta^2 >\frac{1}{4}$ is satisfied and 
$\Lambda$ is greater than $\frac{1}{4}$.
These two cases are complementary to each other 
in view of the values of $\Lambda$. 

Next we turn to the states $\ket{\Lambda,a}$ 
in the first case and investigate properties of the states.
The semi-positivity condition for the operator $J^{\dagger}_{\pm}J_{\pm}$, 
will give allowed regions 
in the $(j,a)$-plane. By recalling the equations, 
\beq
&& J_+J_- \ket{\Lambda ,a}=
(a-j-1)(a+j)\ket{\Lambda ,a}\,,\\
&&
J_-J_+ \ket{\Lambda ,a}=
(a+j+1)(a-j)\ket{\Lambda ,a}\,,
\eeq
we can write down constraints for $(j,a)$,
\beq
&& 0\leq \Lambda +a(a\mp 1)=-j(j+1)+a(a\mp 1)\,,\\
&&(a-j-1)(a+j)\geq 0 \,\,\mbox{and}\,\,
(a+j+1)(a-j)\geq 0\, .
\eeq
By recalling $j<0$, we have a condition $a\geq -j$ for
a state $\ket{\Lambda ,a_0}$. 
If $a>-j$ and $a\neq -j$ are satisfied, 
we will obtain non suitable states 
by acting $J_{-}$ to the state $\ket{\Lambda ,a_0}$ repeatedly. 
So we need a condition $a_0=-j$ for an appropriate $a=a_0$ and 
obtain a discrete series of the states $D^+(j)$ which are 
constructed 
from $\ket{\Lambda ,a_0}$ by applying $J_+$ repeatedly. 
The eigenvalues $a$'s of the states in this series
are bounded from below and 
written as $a=a_0, a_0+1 ,a_0+2,\cdots$.

As the second case, we have a discrete series $D^{-}(j)$ of the states 
by considering a condition $a<j$ $(j<0)$. 
The states in this series are labelled by 
$a$ and $j$. 
In this series, the $a$'s take their values in discrete numbers $a=a_0,a_0-1, a_0-2,\cdots$ 
which are bounded from above. 

As the third case, we have continuous supplementary series $D_s(\Lambda ,a_0)$. 
The states in this representtion have eigenvalues $a$'s 
that take their values in discrete numbers 
$a=a_0, a_0\pm 1,a_o\pm 2,\cdots$. They are neither bounded from above nor 
from below.

We have another continuous representation 
$D_p(\Lambda ,a_0)$ labelled by 
a complex number 
$j=-\frac{1}{2}+i\beta$ $(\beta \in {\bf R})$ 
that is called as the continuous principal series. 
The states are characterised by the eigenvalues 
$a$'s and their values are writted as 
$a=a_0, a_0\pm 1, a_0\pm 2,\cdots$. 

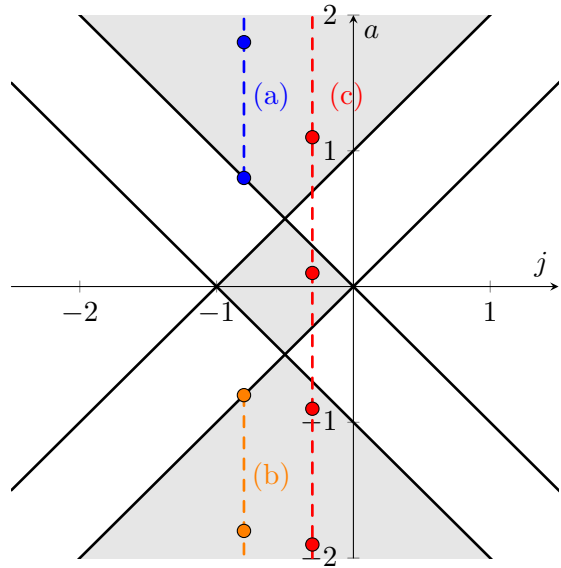
\begin{figure}
\centering
\definecolor{ffxfqq}{rgb}{1,0.5,0}
\definecolor{qqqqff}{rgb}{0,0,1}
\definecolor{ffqqqq}{rgb}{1,0,0}
\begin{tikzpicture}[line cap=round,line join=round,>=triangle 45,x=1cm,y=1cm]
\begin{axis}[
x=1.8cm,y=1.8cm,
axis lines=middle,
xmin=-2.5,
xmax=1.5,
ymin=-2.0,
ymax=2.0,
xtick={-2,-1,...,1},
ytick={-2,-1,...,2},
xlabel=$j$,
ylabel=$a$
]
\clip(-2.5,-2.0) rectangle (2.0,2.0);
\fill[line width=2pt,fill=black,fill opacity=0.1] (-0.5,0.5) -- (2,3) -- (-3,3) -- cycle;
\fill[line width=2pt,fill=black,fill opacity=0.1] (-0.5,-0.5) -- (0,0) -- (-0.5,0.5) -- (-1,0) -- cycle;
\fill[line width=2pt,fill=black,fill opacity=0.1] (-0.5,-0.5) -- (2,-3) -- (-3,-3) -- cycle;
\draw [line width=1pt,domain=-2.5:2.0] plot(\x,{(-0--1*\x)/-1});
\draw [line width=1pt,domain=-2.5:2.0] plot(\x,{(--1--1*\x)/1});
\draw [line width=1pt,domain=-2.5:2.0] plot(\x,{(-1-1*\x)/1});
\draw [line width=1pt,domain=-2.5:2.0] plot(\x,{(-0-1*\x)/-1});
\draw [line width=1pt,dash pattern=on 4pt off 4pt,color=qqqqff] (-0.8,0.8) -- (-0.8,2.0);
\draw [line width=1pt,dash pattern=on 4pt off 4pt,color=ffxfqq] (-0.8,-0.8) -- (-0.8,-2.0);
\draw [line width=1pt,dash pattern=on 4pt off 4pt,color=ffqqqq] (-0.3,-2.0) -- (-0.3,2.0);
\begin{scriptsize}
\draw [fill=ffqqqq] (-0.3,0.1) circle (2.5pt);
\draw [fill=ffqqqq] (-0.3,1.1) circle (2.5pt);
\draw [fill=ffqqqq] (-0.3,-0.9) circle (2.5pt);
\draw [fill=qqqqff] (-0.8,0.8) circle (2.5pt);
\draw [fill=qqqqff] (-0.8,1.8) circle (2.5pt);
\draw [fill=ffxfqq] (-0.8,-0.8) circle (2.5pt);
\draw [fill=ffxfqq] (-0.8,-1.8) circle (2.5pt);
\draw [fill=ffqqqq] (-0.3,-1.9) circle (2.5pt);
\draw[color=qqqqff] (-0.6,1.4) node {(a)};
\draw[color=ffxfqq] (-0.6,-1.4) node {(b)};
\draw[color=ffqqqq] (-0.05,1.4) node {(c)};
\draw [fill=ffqqqq] (-0.3,2.1) circle (2.5pt);
\draw [fill=ffqqqq] (-0.3,-2.9) circle (2.5pt);
\draw [fill=qqqqff] (-0.8,2.80) circle (2.5pt);
\draw [fill=ffxfqq] (-0.8,-2.8) circle (2.5pt);
\end{scriptsize}
\end{axis}
\end{tikzpicture}
\caption{
    Representations of $su(1,1)$.
    The allowed regions for a combination $(j,a)$ are illustrated by shaded regions.
    The blue blobs belong to (a) discrete series bounded from below,
    the orange ones belong to (b) discrete series bounded from above,
    and the red ones belong to (c) continuous supplementary series which is unbounded.
}
\label{fig:repr_su11}
\end{figure}

We summarize these four representations in the following.
The allowed regions are illustrated by shaded regions in the Fig.~\ref{fig:repr_su11}.
There are four types of representations. Three of them are associated to 
these regions in the Fig.~\ref{fig:repr_su11}. 
The remaining one corresponds to the case of the complex 
$j=-\frac{1}{2}+i\beta $ $(\beta\in {\bf R})$. 
The states are represented by $\ket{\Lambda,a}$ with 
$\Lambda  =-j(j+1)$:
\begin{enumerate}
\item[] (a) Discrete series $D^+(j)$; bounded from below
$\cdots J_- \ket{\Lambda ,a_0}=0$
\beq
&& a_0:=-j\quad (j<0)\\
&& a=a_0, a_0+1, a_0+2,\cdots 
\eeq

\item[] (b) Discrete series $D^-(j)$; bounded from above 
$\cdots J_+ \ket{\Lambda ,a_0}=0$
\beq
&& a_0:=j\quad (j<0)\\
&& a=a_0, a_0-1, a_0-2,\cdots 
\eeq

\item[] (c) Continuous supplementary series $D_s(\Lambda ,a_0)$ 
; unbounded
\beq
&& \left|j+\frac{1}{2} \right|<\frac{1}{2}-|a_0|\,,\\
&&  a=a_0, a_0\pm 1, a_0\pm 2,\cdots 
\eeq

\item[] (d) Continuous principal series $D_p(\Lambda ,a_0)$
; unbounded
\beq
&& j=-\frac{1}{2} +i\beta \quad (\beta >0)\\
&&  a=a_0, a_0\pm 1, a_0\pm 2,\cdots 
\eeq

\end{enumerate}

 \section{Solvable models and symmetry}
 \label{sec:B}

In this appendix, we summarize the solvable models used in Sec.~\ref{sec:4} and discuss their spectra 
from the viewpoint of the symmetry studied in App.~\ref{sec:A}.

\subsection{Inverted harmonic oscillator}
 \label{sec:B-1}

\subsubsection{\texorpdfstring{$ SL(2;{\mathbf R})$}{TEXT} symmetry and spectra}
\label{sec:B-1-1}

Using the representation theory described in Sec.~\ref{sec:A}, we can obtain the spectra of a harmonic oscillator or an inverted harmonic oscillator. This is because the Hamiltonians of these systems in fact satisfy the algebra.

Let us consider a set of differential operators 
$J_a$'s $(a=1,2,3)$ as  an example of 
a realization of this symmetry algebra $su(1,1)$:
\beq
&& J_1=\frac{1}{2} \left(\frac{d^2}{dx^2} +\frac{\tilde{a}}{x^2}+\frac{x^2}{4}
\right)\,,\\
&& J_2=\frac{-i}{2}\left(x\frac{d}{dx} +\frac{1}{2}\right)\,,
\label{J_2}
\\
&& J_3=\frac{1}{2} \left(\frac{d^2}{dx^2} 
+\frac{\tilde{a}}{x^2}-\frac{x^2}{4}
\right)\,,\\
&& (J_a)^{\dagger}=J_a\quad (a=1,2,3)\,.
\eeq
The Casimir invariant is constant in this model, which is 
specified by the parameter $\tilde{a}$.
\beq
&& J^2=J_1^2+J_2^2-J_3^2=\frac{\tilde{a}}{4}+\frac{3}{16}=
-j(j+1)\,.
\eeq
Then $j$ is expressed  by using $\tilde{a}$ as
$ j=-\frac{1}{2}\pm \frac{1}{2}
\sqrt{\frac{1}{4}-\tilde{a}}$.

Now 
we shall study the hidden conformal symmetry proposed in \cite{Raffaelli:2021gzh}
by using 
these differential operators with $\tilde{a}=0$. 
We follow the notation of \cite{Raffaelli:2021gzh}
and redefine the currents as
\beq
&& \hat{J}_3=-i J_1=-\frac{i}{2}\left(\frac{d^2}{dx^2}+\frac{x^2}{4}\right)
=\frac{i}{2}H\,,\\
&&
\hat{J}_1=J_2=-\frac{i}{2}\left(x\frac{d}{dx}+\frac{1}{2}\right)\,,\\
&&
\hat{J}_2=iJ_3=\frac{i}{2}\left(\frac{d^2}{dx^2}-\frac{x^2}{4}\right)
=-\frac{i}{2}H_{\rm ho}\,,\\
&& \hat{J}_{\pm}=\pm i\hat{J}_1-\hat{J}_2 =\pm i J_2-iJ_3\,,\\
&& [\hat{J}_+,\hat{J}_-]=-2 \hat{J}_3\,,\,\,
[\hat{J}_3, \hat{J}_{\pm}]=\pm \hat{J}_{\pm}\,.
\eeq

This system is the inverted harmonic oscillator in 
1 dimension with the Hamiltonian $H$. 
The spectrum of this model is represented as 
eigenvalues of $H=-2i \hat{J}_3$.
In addition, the Casimir invariant is defined as
\beq
&& \hat{J}^2=\hat{J}_3^2-\hat{J}_1^2-\hat{J}_2^2
=-J^2\,,\\
&&\hat{J}^2=\hat{J}_3^2-\frac{1}{2}(\hat{J}_+\hat{J}_-+\hat{J}_-
\hat{J}_+)=\hat{J}_3^2\pm \hat{J}_3-\hat{J}_{\mp}\hat{J}_{\pm}\,,\\
&&J^2 \cong \Lambda =-j(j+1)\,,\,\,\\
&&
\hat{J}^2\cong \hat{\Lambda}=-\Lambda =j(j+1)=\hat{j}(\hat{j}-1)\,,\,\,\\
&& \hat{j}:=j+1\,.
\eeq

The states are labelled by a set of integers $(\hat{j}, m)$ with
\beq
&& \hat{J}^2\cong \hat{\Lambda} =\hat{j}(\hat{j}-1)\,,\\
&& \hat{J}_3\cong m\,.
\eeq
By using the explicit form of the differential operators, 
one can obtain a relation
\beq
\hat{J}^2=-\frac{3}{16}\,.
\eeq
It determines the values of $\hat{j}$ as $\hat{j}=\frac{1}{4}, \frac{3}{4}$. 
We introduce a useful notation to label these two eigenvalues  as
\beq
&&
\hat{j}_{\alpha}=\frac{2\alpha -1}{4}\quad (\alpha =1,2)\,,\\
&&
\hat{J}^2 \cong \hat{j}_{\alpha} (\hat{j}_{\alpha} -1)\,,\\
&& \hat{J}_3 \cong m\,.
\eeq

We can change the eigenvalue $m$ by a unit by 
acting the raising $\hat{J}_+$ and the lowering $\hat{J}_-$ operators 
on the state, to find
\beq
&& \hat{J}_{\mp}\hat{J}_{\pm} \ket{\hat{j} ,m}
=(m\pm \hat{j})\{m\pm (1-\hat{j})\} \ket{\hat{j} ,m}\nonumber \\
&&
=(m\pm \frac{1}{4})(m\pm \frac{3}{4}) \ket{\hat{j} ,m}\,.
\eeq

Let us take the following two representations of this algebra and 
estimate the energy spectrum $H=-2i \hat{J}_3$.
\begin{enumerate}
\item Spectrum bounded from below
\beq
&& \hat{J}_- \ket{\hat{j}_{\alpha} ,m_0}=0\cdots 
m_0=+\hat{j}_{\alpha}\,,\\
&& \hat{J}_+^k \ket{\hat{j}_{\alpha} ,+\hat{j}_{\alpha}} 
\cdots H\cong -2i (k+\hat{j}_{\alpha})=
-i(n+\frac{1}{2})\,,\\
&& n=2k+\alpha -1\,.
\eeq

\item Spectrum bounded from above
\beq
&& \hat{J}_+ \ket{\hat{j}_{\alpha} ,m_0}=0\cdots 
m_0=-\hat{j}_{\alpha}\,,\\
&& \hat{J}_-^k \ket{\hat{j}_{\alpha} ,-\hat{j}_{\alpha}} 
\cdots H\cong -2i (-k-\hat{j}_{\alpha})=
+i(n+\frac{1}{2})\,,\\
&& n=2k+\alpha -1\,.
\eeq
\end{enumerate}

Then one can construct eigenstates $ \ket{n,\pm}$ $(n=0,1,2,\cdots )$ and 
associated eigenfunctions $\tilde{\Phi}^{(\pm)}_n (x)$ as
\beq
&& \ket{n,\pm}:=\hat{J}^k_{\pm}\ket{\hat{j}_{\alpha}, \pm \hat{j}_{\alpha}}\,,\\
&& n=2k+\alpha -1\,,\,\, \hat{j}_{\alpha} =\frac{2\alpha -1}{4}\qquad (\alpha =1,2)\,,
\eeq
\beq
&& \tilde{\Phi}^{(\pm)}_n (x)=\langle x| n, \pm  \rangle \,,\\
&& \tilde{\Phi}^{(\pm)}_{0,\alpha}=x^{\alpha -1}\cdot e^{\pm \frac{i}{4}x^2}
\quad (\alpha =1,2)\,,\\
&& \tilde{\Phi}^{(\pm)}_{k,\alpha}(x)=e^{\pm \frac{i}{4} x^2}\cdot
H_n \left(e^{\mp \frac{\pi i}{4}}\frac{x}{\sqrt{2}}\right)\,,\\
&& H_n(x)=(-1)^n e^{x^2}\left(\frac{d}{dx}\right)^n e^{-x^2}\,.
\eeq
This system is the inverted harmonic oscillator, and 
its spectra match the eigenvalues of the Schr\"odinger equation.
We will see the details of the wave functions in Sec.~\ref{sec:B-1-2}.

The spectra are also related to those of the usual harmonic oscillator by a change of variables: 
by considering a rotation in the $2,3$-plane, we can transform 
the equation of the harmonic oscillator $(H_{\rm ho}\tilde{\Phi}_{\rm ho}=E_{\rm ho}\tilde{\Phi}_{\rm ho})$ into 
that of the inverted potential case $(H \tilde{\Phi} =E\Phi)$, 
with $D_{\pm}:=\exp[\pm \frac{\pi}{2}\hat{J}_1]$, as
\beq
&&D_{\pm} \cdot H\cdot D^{-1}_{\pm} =\pm i H_{\rm ho}\,,\\
&&\tilde{\Phi}(x)=D_{\pm}^{-1}\tilde{\Phi}_{\rm ho}(x)
=e^{\pm \frac{\pi i }{8}}\tilde{\Phi}_{\rm ho}(e^{\pm \frac{\pi i}{4}}x)\,,\\
&& E=\pm i E_{\rm ho}\, .
\eeq

 \subsubsection{Wave functions and spectra}
 \label{sec:B-1-2}

Here let us turn to the Schr\"odinger equation of
the inverted harmonic oscillator in one dimension, and look at the solutions
to identify what kind of boundary conditions are implicitly imposed in the
spectrum obtained by the symmetry argument in Sec.~\ref{sec:B-1-1}.

 The Schr\"odinger equation determines the energy eigenvalue $E$ and 
 the associated wavefunction $\psi (x)$
\beq
&&
\left(\frac{d^2}{dx^2}+E+\frac{1}{4}x^2\right)\psi =0\,.
\eeq
By changing variables $z:=e^{\frac{\pi i}{4}}\cdot x$, 
$\nu :=-\frac{1}{2}-i E$, one can rewrite this equation into 
a differential equation 
\beq
&&
\left(\frac{d^2}{dz^2}+\nu +\frac{1}{2}-\frac{1}{4}z^2\right)\psi =0\,.
\label{diffeqiho}
\eeq
This has two independent  solutions expressed by the parabolic cylinder functions
$D_{\nu}(z)$, $D_{-\nu -1}(i z)$ and 
$\psi (x)$ is described as a linear combination of them 
\beq
\psi (z)=A \cdot D_{\nu}(z)+B\cdot D_{-\nu -1}(iz)\,.
\eeq

We here summarize  several useful formulas of the 
parabolic cylinder function.
The function $D_{\nu}(z)$ is  represented by 
the confluent hypergeometric function
\beq
&&
D_{\nu}(z)=2^{\frac{\nu}{2}}\sqrt{\pi}\cdot 
e^{-\frac{z^2}{4}}
\Biggl[
\frac{1}{\Gamma (\frac{1-\nu}{2})} 
F \!\left( \begin{array}{c} -\frac{\nu}{2} \\ \frac{1}{2} \\ \end{array} ; 
\frac{z^2}{2}\right)
-\frac{\sqrt{2}}{\Gamma (-\frac{\nu}{2})} z \cdot 
F \left( \begin{array}{c} \frac{1-\nu}{2} \\ \frac{3}{2} \end{array} ;\frac{z^2}{2}\right)
\Biggr]\,.
\eeq
The differential equation \eqref{diffeqiho}
has four types of solutions
$D_{\nu}(z)$,
$D_{-\nu}(z)$,
$D_{-\nu -1}(iz)$, and
$D_{-\nu -1}(-iz)$. 
They are related one another by the following identities
\beq
&&
D_{\nu}(z)=
\frac{\Gamma (\nu +1)}{\sqrt{2\pi }}
\left[
e^{+\frac{\pi i}{2}\nu}D_{-\nu -1}(iz)
+e^{-\frac{\pi i}{2}\nu}D_{-\nu -1}(-iz)
\right]\,,\\
&&
D_{\nu}(z)=
e^{\pm \pi i\nu}D_{\nu}(-z)
+\frac{\sqrt{2\pi }}{\Gamma (-\nu)}
e^{\pm \frac{\pi i}{2}(\nu +1)} D_{-\nu -1}
(\mp iz)\,.
\eeq
The function $D_{\nu}(z)$ has an 
asymptotic form in the region $|z|\gg 1$ $(\nu \neq 0,1,2,\cdots)$ according to
 the argument of $z$,
\beq
D_{\nu}(z) &\sim 
e^{-\frac{z^2}{4}}z^{\nu} -\frac{\sqrt{2\pi}}{\Gamma (-\nu)}
e^{+\nu \pi i } e^{+\frac{z^2}{4}}z^{-\nu -1} \,\,\,
&\left(  \frac{\pi}{4}  < \arg z < \frac{5\pi}{4}\right)\,,\\
D_{\nu}(z) &\sim 
e^{-\frac{z^2}{4}}z^{\nu} -\frac{\sqrt{2\pi}}{\Gamma (-\nu)}
e^{-\nu \pi i } e^{+\frac{z^2}{4}}z^{-\nu -1} \,\,\,
&\left(  -\frac{5 \pi}{4}  < \arg z < -\frac{\pi}{4}\right)\,,\\
D_{\nu}(z) &\sim 
e^{-\frac{z^2}{4}}z^{\nu} \,\,\,
&\left(  -\frac{3\pi}{4}  < \arg z < \frac{3\pi}{4}\right)\,.
\eeq
For $\nu =n\in {\bf Z}_{\geq 0}$, 
$D_{\nu}(z)$ is represented by the Hermite polynomial 
$H_n(z)$
\beq
&& D_n(z)=
(-1)^n e^{\frac{z^2}{4}}\frac{d^n}{dz^n} e^{-\frac{z^2}{2}}
=2^{-\frac{n}{2}} e^{-\frac{z^2}{4}}H_n(\frac{z}{\sqrt{2}}) \,,\\
&&
H_n(z):=(-1)^n e^{z^2}\frac{d^n}{dz^n} e^{-z^2}\,.
\eeq

When there is the inverted harmonic potential, 
the waves are scattered by this repulsive potential. 
In the region $x>0$  $(z=e^{\frac{\pi i}{4}}\cdot x )$, 
the wavefunction $\psi (x)$ behaves at $x \rightarrow +\infty$ as
\beq
&&
\psi (z)\sim A\cdot e^{-\frac{z^2}{4}}\cdot z^{\nu}
+B\cdot e^{+\frac{z^2}{4}}\cdot (iz)^{-\nu -1}\,.
\eeq
If the system time-evolves according to the factor 
$e^{-i\Omega t}$ in the wave function,  
one can see that 
the first term $e^{-\frac{z^2}{4}}=e^{-i\frac{x^2}{4}}$
is related to 
a left-moving wave (in-going mode as seen from the potential hill). 
On the other hand, 
the second term 
$e^{+\frac{z^2}{4}}=e^{+i\frac{x^2}{4}}$ represents a 
right-moving wave (in-going mode). 

In order to analyze the scattering process, 
we want to know the behavior in the region $x\rightarrow -\infty$. 
The two solutions can be transformed by analytic continuations
with  $\tilde{x}:=-x$  $(\tilde{x}>0)$
\beq
&&
\tilde{z}:=e^{\frac{\pi i}{4}}\cdot \tilde{x}\,,\\
&&
\psi (z)=\psi (-\tilde{z})
=A\cdot D_{\nu}(-\tilde{z})+B\cdot D_{-\nu -1}(-i\tilde{z})\\
&&
=A'\cdot D_{\nu}(\tilde{z})+B'\cdot D_{-\nu -1}(i\tilde{z})\,.
\eeq
Now we have two sets of coefficients $(A,B)$ and $(A',B')$ those 
encode the information of  the scattering. 
There are relations between $(A,B)$ and $(A',B')$ :
\beq
&&
\left(
\begin{array}{c}
A' \\
B' \\
\end{array}
\right)
=\left(
\begin{array}{cc}
T_{11}  & T_{12} \\
T_{21} & T_{22} \\
\end{array}
\right)
\left(
\begin{array}{c}
A \\
B \\
\end{array}
\right)\,,\\
&&
T_{11}=e^{\pi i \nu}\,,\quad
T_{21}=\frac{\sqrt{2\pi }}{\Gamma (-\nu)} e^{\frac{\pi i}{2}(\nu +1)}\,,
T_{12}=\frac{\sqrt{2\pi }}{\Gamma (\nu +1)} e^{\frac{\pi i}{2}\nu}\,,\,\,
T_{22}=e^{\pi i (\nu +1)}\,.
\eeq
In the region $x \to -\infty$, 
the first term $e^{-\frac{z^2}{4}}=e^{-i\frac{x^2}{4}}$
with $A'$
is a right-moving wave (in-going mode), while 
the second term 
$e^{+\frac{z^2}{4}}=e^{+i\frac{x^2}{4}}$ with $B'$
represents a left-moving wave (out-going mode). 

The boundary conditions which one likes to impose for the wave functions are
used for determining the coefficients $(A, B)$ and $(A', B')$.
To illustrate how the condition on the coefficients can determine the spectrum,
let us consider quasinormal modes
with the condition 
\beq
A=A'=0\, .
\eeq
This condition amounts to the boundary condition of having purely out-going modes
on the both sides of the potential hill.
We have an equation  
$T_{12}=\frac{\sqrt{2\pi }}{\Gamma (\nu +1)} e^{\frac{\pi i}{2}\nu}=0$ 
from 
$A'=T_{11}A+T_{12}B=T_{12}B=0$. 
The poles of  $\Gamma (\nu +1)$ have contribution to this equation
and we have a series of solutions 
\beq
\nu =-n-1 \qquad (n=0,1,2,\cdots )\,.
\label{nun1}
\eeq
 In this discrete series, 
 the associated wavefunction can be described by 
 the Hermite polynomial
\beq
&&
\psi (z)=B\cdot D_{-\nu -1}(iz)=B\cdot D_{n}(iz)\,,\\
&&
D_n (z)=(-1)^n e^{\frac{z^2}{4}}\frac{d^n}{dz^n} e^{-\frac{z^2}{2}}
=2^{-\frac{n}{2}}\cdot e^{-\frac{z^2}{4}} H_n(\frac{z}{\sqrt{2}})\,,\\
&&
H_n(z)=(-1)^n e^{z^2}\frac{d^n}{dz^n} e^{- z^2 }\,.
\eeq

Now we shall relate this spectral formula \eqref{nun1} to the quasinormal mode spectrum of the black hole. 
When one considers a wave equation of the scalar field $\Phi$, 
 the radial part $\psi (r)$ is described by the Schr\"odinger equation
\beq
&&
\left(\frac{d^2}{d r_{\ast}^2}+\Omega^2-V(r)\right)\psi =0\,,\\
&&
V=V(r_\mathrm{PS})+\frac{1}{2}(f(r_\mathrm{PS})^2\cdot V''(r_\mathrm{PS}))\cdot (\delta {r}_{\ast})^2\,,
\,\,\\
&&
\delta {r}_{\ast}=r_{\ast}-(r_{\ast})_\mathrm{PS} \,.
\eeq
When we associate $x$ and $E$ with $\delta {r}_{\ast}$ and $\Omega$, 
this system is reduced to the inverted harmonic oscillator 
\beq
&&
x=(-2\cdot f(r_\mathrm{PS})^2\cdot V''(r_\mathrm{PS}))^{1/4}\cdot \delta {r}_{\ast}\,,\\
&&
E=\frac{\Omega^2 -V(r_\mathrm{PS})}{(-2\cdot f(r_\mathrm{PS})^2\cdot V''(r_\mathrm{PS}))^{1/2}}\,,\\
&&
\left(\frac{d^2}{dx^2}+E+\frac{1}{4}x^2\right)\psi =0\,.
\eeq
By recalling the above analysis, we can write down 
the values $E$ and $\Omega^2$ that are 
characterised by a set of integer $n$ $(n=0,1,2,\cdots)$
\beq
&&
E=i \left(\nu +\frac{1}{2}\right)=-i \left(n+\frac{1}{2}\right)\quad (n=0,1,2,\cdots)\,,\\
&&
\Omega^2=V(r_\mathrm{PS})+(-2\cdot f(r_\mathrm{PS}))^2\cdot V''(r_\mathrm{PS}))^{1/2}\times E \\
&&
\qquad =V(r_\mathrm{PS})+2\sqrt{V(r_\mathrm{PS})}\cdot \gamma_L \times (-i)(n+\frac{1}{2}) \,.
\eeq
If there is a conditon $\frac{2\gamma_L}{\sqrt{V(r_\mathrm{PS})}}\left(n+\frac{1}{2}\right)\ll1$, 
we have a  formula for the quasinormal mode spectrum $\Omega$ as 
\beq
&&
\Omega =\sqrt{V(r_\mathrm{PS})}-i \gamma_L \cdot \left(n+\frac{1}{2}\right)\,.
\eeq
In the case of the Schwarzschild black hole, 
one can estimate
$\sqrt{V(r_\mathrm{PS})}=\ell \frac{\sqrt{f(r_\mathrm{PS})}}{r_\mathrm{PS}}$, $\gamma_L=\frac{\sqrt{f(r_\mathrm{PS})}}{r_\mathrm{PS}}$. 
Then we obtain a restriction
$\frac{2}{\ell}\times (n+\frac{1}{2})\ll1$.
This is the spectrum obtained in \cite{Hadar:2022xag}, and we conclude that the spectrum generated by the symmetry
amounts to the boundary condition of purely out-going modes on the both sides of the potential hill.

In our case of the AdS Schwarzschild black hole studeied in this paper, 
the existence of the hard wall of the AdS boundary actually
breaks the $SL(2;\mathbf{R})$ symmetry. This can be seen by one of the the symmetry generators \eqref{J_2} which is a scaling operator. A concrete choice of the position of the hard wall $x=L$ spoils this generator, that is why our quasinormal mode spectrum in the AdS Schwarzschild black hole does not follow the $SL(2;\mathbf{R})$ symmetry.

\subsection{P\"oschl-Teller model}
\label{sec:B-2}

 We summarise properties of the P\"oschl-Teller model and 
 the related algebra studied in the paper 
 \cite{Cevik:2016mnr}.

\subsubsection{Wave functions and spectra}
 
 The model is the integrable system in one dimension 
 with a potential $V(x)=\frac{V_0}{\cosh^2 \alpha x}$,  
 $V_0=-\alpha^2 \lambda (\lambda -1)$. The parameter $\alpha >0$ characterises the
 width of the potential in the $x$-direction and $\lambda$ specifies the 
 height of the peak $V_0$ of the potential.  In this paper, we consider 
 the repulsive case with $V_0 >0$ and study the scattering processes by the barrier 
 generated by $V(x)$. 
 The dynamics is described by a Hamiltonian $H$. 
 The energy $\omega^2$ and the eigenfunction $\psi (x)$ are determined by a Schr\"odinger equation 
 \beq
&& V=\frac{V_0}{\cosh^2 \alpha x}\,,\,\,
V_0=-\alpha^2 \lambda (\lambda -1)\,,\\
&&
H=-\frac{d^2}{dx^2}+V(x)\,,\,\,\\
&& H\psi =\omega^2 \psi\,.
\eeq
The wavefunction is expressed as a linear combination of 
two independent solutions $f_{\pm}(x)$
\beq
&&
\psi (x)=A\cdot  f_+ (x)+B\cdot  f_- (x)\,,\\
&&
f_+(x)
=(e^{ \alpha x})^{\frac{i\omega}{\alpha}}\times 
\geom{\lambda}{1-\lambda}{1+\frac{i\omega}{\alpha}}
{\frac{e^{\alpha x}}{e^{\alpha x}+e^{-\alpha x}}}\,,\\
&&
f_-(x)
=(e^{\alpha x}+e^{-\alpha x})^{\frac{i\omega}{\alpha}}
\times 
\geom{\lambda -\frac{i\omega}{\alpha}}
{1-\lambda -\frac{i\omega}{\alpha}}{1-\frac{i\omega}{\alpha}}
{\frac{e^{\alpha x}}{e^{\alpha x}+e^{-\alpha x}}}\,.
\eeq
In the large negative region $x\rightarrow -\infty$, 
we can evaluate the asymptotic bahavior of $\psi (x)$ 
\beq
&&
f_+(x)\sim e^{i\omega x}\,,\,\,
f_-  (x)\sim e^{-i\omega x}\,,\\
&&
\psi (x)\sim A e^{i\omega x}+Be^{-i\omega x}\,.
\eeq
In order to analyse the scattering, we want to know an asymptotic behavior of $\psi (x)$ 
in the large positive region $x\rightarrow +\infty$. 
The two solutions $f_{\pm}(x)$ can be rewritten by analytic continuations
\beq
&&
f_+(x)=
e^{+i\omega x}\cdot T_{11}(\omega )
\times 
\geom{\lambda} {1-\lambda} {1-\frac{i\omega}{\alpha}}
{\frac{e^{-\alpha x}}{e^{\alpha x}+e^{-\alpha x}}}\nonumber \\
&&
\qquad +
(e^{\alpha x}+e^{-\alpha x})^{-\frac{i\omega}{\alpha}}\cdot 
T_{21}(\omega )
\times 
\geom{\lambda +\frac{i\omega}{\alpha}}
{1-\lambda +\frac{i\omega}{\alpha}}{1+\frac{i\omega}{\alpha}}
{\frac{e^{-\alpha x}}{e^{\alpha x}+e^{-\alpha x}}}\,,\\
&&
f_-(x)=
e^{-i\omega x}\cdot T_{22}(\omega )
\times 
\geom{\lambda} {1-\lambda} {1+\frac{i\omega}{\alpha}}
{\frac{e^{-\alpha x}}{e^{\alpha x}+e^{-\alpha x}}} \nonumber \\
&&
\qquad +
(e^{\alpha x}+e^{-\alpha x})^{+\frac{i\omega}{\alpha}}\cdot 
T_{12}(\omega )
\times 
\geom{\lambda -\frac{i\omega}{\alpha}}
{1-\lambda -\frac{i\omega}{\alpha}}{1-\frac{i\omega}{\alpha}}
{\frac{e^{-\alpha x}}{e^{\alpha x}+e^{-\alpha x}}}\,,
\eeq
where coefficients $T_{11}$, $T_{12}$, $T_{21}$, $T_{22}$ are defined as
\beq
&&
T_{11}=\frac{\Gamma (1+\frac{i\omega}{\alpha}) \Gamma (\frac{i\omega}{\alpha})}
{\Gamma (1-\lambda +\frac{i\omega}{\alpha}) \Gamma (\lambda +\frac{i\omega}{\alpha})}\,,\\
&&
T_{21}=\frac{\Gamma (1+\frac{i\omega}{\alpha}) \Gamma (-\frac{i\omega}{\alpha})}
{\Gamma (1-\lambda ) \Gamma (\lambda )}\,,\\
&&
T_{12}=\frac{\Gamma (1-\frac{i\omega}{\alpha}) \Gamma (\frac{i\omega}{\alpha})}
{\Gamma (1-\lambda ) \Gamma (\lambda )}\,,\\
&&
T_{22}=\frac{\Gamma (1-\frac{i\omega}{\alpha}) \Gamma (-\frac{i\omega}{\alpha})}
{\Gamma (1-\lambda -\frac{i\omega}{\alpha}) \Gamma (\lambda -\frac{i\omega}{\alpha})}\,.
\eeq
In the region $x\rightarrow +\infty$, 
the wavefunction $\psi (x)$ behaves as
\beq
&&
f_+(x)\sim T_{11}(\omega)e^{i\omega x}+T_{21}(\omega)e^{-i\omega x}\,,\,\,\\
&&
f_-  (x)\sim 
T_{12}(\omega)e^{i\omega x}+T_{22}(\omega)e^{-i\omega x}\,,\,\,\\
&&
\psi (x) \sim A' e^{i\omega x}+B' e^{-i\omega x}\,,
\eeq
where $A'$, $B'$ are related to $A$, $B$ by equations
\beq
&&
\left(
\begin{array}{c}
A' \\
B' \\
\end{array}
\right)
=
\left(
\begin{array}{cc}
T_{11} & T_{12} \\
T_{21} & T_{22} \\
\end{array}
\right)
\left(
\begin{array}{c}
A \\
B \\
\end{array}
\right)\,.
\eeq
When we take the point $x=0$ as a reference point of the scattering, 
the waves with coefficients $A$, $B'$ are 
in-going ones. On the otherhand, $B$ and $A'$ counterparts 
are out-going waves.  
As boundary conditions, we put $A=B'=0$ and obtain 
a relation $T_{22}(\omega)
=\frac{\Gamma (1-\frac{i\omega}{\alpha}) \Gamma (-\frac{i\omega}{\alpha})}
{\Gamma (1-\lambda -\frac{i\omega}{\alpha}) \Gamma (\lambda -\frac{i\omega}{\alpha})}
=0$. The gamma functions in the denominator have poles 
in the $\omega$-plane and they
determine quasinormal modes $\omega^{(\pm)}_n$ $(n=0,1,2,\cdots) $
\beq
&&
\omega^{(\pm)}_n= -i\alpha \left( n+\lambda^{(\pm)}\right) \,,\,\,
\qquad (n=0,1,2,\cdots)\\
&&
\lambda^{(+)}:=\lambda \,,\,\,
\lambda^{(-)}:=1-\lambda\,.
\eeq
In this case,  the wavefunction $\psi (x)=B\cdot \psi^{(\pm)}_n(x)$ is written by 
the hypergeometric function
\beq
&&
\psi^{(\pm)}_n(x)=
(e^{\alpha x}+e^{-\alpha x})^{\frac{i\omega_n^{(\pm)}}{\alpha}}
\nonumber \\
&&
\qquad \times 
\geom{\lambda -\frac{i\omega^{(\pm)}_n}{\alpha}} 
{1-\lambda -\frac{i\omega^{(\pm)}_n}{\alpha}}
{1-\frac{i\omega^{(\pm)}_n}{\alpha}} {\frac{e^{\alpha x}}{e^{\alpha x} +e^{-\alpha x}}}
\,\nonumber \\
&& \qquad  
=
(e^{\alpha x}+e^{-\alpha x})^{n +\lambda^{(\pm)}}
\nonumber \\
&&
\qquad \times 
\geom{-n} {1-2 \lambda^{(\pm)} -n}  
{1-\lambda^{(\pm)} -n} {\frac{e^{\alpha x}}{e^{\alpha x} +e^{-\alpha x}}}
\,.
\eeq
 
 Now 
 we will put $\lambda =\frac{1}{2}+i \nu$ $(\nu \in {\bf R})$ 
 to reduce this potential model to the 
 quasinormal modes of the black hole. It is the 
 high barrier case and is related to the continuous representation of the 
 Lie algebra $sl(2;{\bf R})$.
 In that case,  the peak of $V(x)$ is 
 $V_0=\alpha^2 (\frac{1}{4}+\nu^2)$ which is labelled by $\nu \in {\bf R}$ 
 and the modes $\omega_n^{(\pm)}$'s are labelled by $\nu$
\beq
&&
\omega^{(\pm)}_n=-i\alpha \left(n+\frac{1}{2}\right)\pm \alpha \nu \,.
\eeq
The wave propagates  
on the $x$-direction according to $e^{-i\omega t}$.
The amplitue decreases exponentially 
$e^{-i\omega^{(\pm )}_n t}=e^{-\alpha (n+\frac{1}{2})t}e^{\mp i \alpha \nu t}$ 
with $\alpha =\gamma_L$. 
The parameter $\alpha$ is related to the Lyapunov exponent $\gamma_L$ 
and specifies the time scale of the relaxation in the system. 

The associated wavefunctions are also expressed by the Jacobi polynomial
\beq
&&
y=\tanh \alpha x\,,\\
&&
\psi^{(\pm)} _n(x)
=(2\cosh \alpha x)^{n+\frac{1}{2}\pm i\nu}
\times 
\geom{-n} {-n \mp 2i \nu } {-n+\frac{1}{2} \mp i\nu }{\frac{1+y}{2}}
\nonumber \,,\\
&&
\geom{-n} {-n \mp 2i\nu } { -n+\frac{1}{2}\mp i \nu } {\frac{1+y}{2}}
 \nonumber \\
&& \qquad =
\left\{\frac{(-1)^n}{n!}
\frac{\Gamma (\mp i \nu +\frac{1}{2})}{\Gamma (\mp i \nu +\frac{1}{2}-n)}\right\}^{-1}
\times 
P_n^{(\mp i \nu -\frac{1}{2}-n ,\mp i \nu -\frac{1}{2}-n)}(y)\,, \nonumber \\
&&
P_n^{(\mp i \nu -\frac{1}{2}-n ,\mp i \nu -\frac{1}{2}-n)}(y)\nonumber  \\
&& 
\qquad =\frac{(-1)^n}{2^n \cdot n!}\,
(1-y^2)^{n+\frac{1}{2}\pm i\nu}
\left(\frac{d}{dy}\right)^n (1-y^2)^{-\frac{1}{2}\mp i\nu}\,.
\eeq

We shall summarize useful relations for these functions
\beq
&&
P_n^{(\alpha ,\beta)}(x)
=\frac{(-1)^n}{2^n\cdot n!}
(1-x)^{-\alpha}(1+x)^{-\beta}
\left(\frac{d}{dx}\right)^n [(1-x)^{\alpha +n}(1+x)^{\beta +n}]\nonumber \,,\\
&&
P_n^{(\alpha ,\beta)}(x)=
\frac{(-1)^n}{n!}\frac{\Gamma (n+1+\beta)}{\Gamma (1+\beta)}
\times 
\geom{-n} { n+\alpha +\beta +1} {1+\beta } {\frac{1+x}{2}} \nonumber \\
&&
=
\frac{1}{n!}\frac{\Gamma (n+1+\alpha )}{\Gamma (1+\alpha)}
\times 
\geom{-n} {n+\alpha +\beta +1} {1+\alpha }{\frac{1-x}{2}}\,.
\eeq

 We also collect several properties of the hypergeometirc functions here: 
 The hypergeometric equation is defined as
\beq
&& z(1-z)\frac{d^2 f}{dz^2} 
+[c-(a+b+1)z]\frac{df}{dz} -ab f=0\,.
\eeq
There is a set of solutions analytic around  a point $z=0$
\beq
&&
f=\geom{a}{b}{c}{z}\,,\,\,\\
&&
f=z^{1-c}\cdot 
\geom{a-c+1}{b-c +1} {2-c} {z}\,.
\eeq
They are suitable linearly independent solutions that converge under conditons
$|z|<1$, $\mbox{Re}(a+b-c)<0$.

One can reexpress the hypergeometirc function 
defined around $z=0$ in terms of  
analytic functions around $z=1$ through an 
analytic continuation,
\beq
&&
\geom{a}{b}{c}{z}=
\frac{\Gamma (c)\Gamma (a+b-c)}{\Gamma (a)\Gamma (b)}
(1-z)^{c-a-b}
\times \geom{c-a}{c-b}{c-a-b+1}{1-z}\nonumber \\
&& \qquad \qquad \qquad +\frac{\Gamma (c)\Gamma (c-a-b)}{\Gamma (c-a)\Gamma (c-b)}
\times \geom{a}{b}{a+b-c+1}{1-z}\,,\\
&&\qquad\qquad \qquad
\qquad\qquad |\mbox{arg}(1-z)|<\pi\,.\nonumber
\eeq

\subsubsection{\texorpdfstring{$ SL(2;{\mathbf R})$}{TEXT} symmetry and spectra}

 
 The P\"oschl-Teller model can be solved by 
 the Jacobi polynomials in the high barrier case. 
 We will discuss the symmetry of this model proposed in 
  \cite{Cevik:2016mnr}
  in view of the Schr\"odinger equations. 
 Our notation is different from that of 
 \cite{Cevik:2016mnr}. 
 By introducing a variable $y=\alpha x$, we rewrite 
 the Hamiltonian as
 \beq
&&
H=-\frac{d^2}{dy^2}+V(y)\,,\,\, V(y)=-\frac{\lambda (\lambda -1)}{\cosh^2 y}\,,\\
&& H \psi  =\frac{\omega^2}{\alpha^2} \psi \,,
\eeq
and use a new paramater $K_n$ defined as
\beq
&&
K_n=\frac{i}{\alpha}\omega_n^{(\pm)}=
\left\{
\begin{array}{ll}
n+\lambda & (+)\,\, \mbox{case}\\
n+1-\lambda & (-)\,\,\mbox{case} \\
\end{array}
\right.\qquad (n=0,1,2,\cdots )\,.\nonumber 
\eeq
From now on, we abbreviate the superscripts $(\pm)$ 
and introduce a set of operators $J^{\pm}$, $J^3$ 
that act on the solutions $\{\psi_n\}$ $(n=0,1,2,\cdots)$
\beq
&&
J^-_n\,;\,\psi_n\rightarrow \psi_{n-1}\,,\,\,
J^+_{n+1}\,;\,\psi_{n}\rightarrow \psi_{n+1}\,,\\
&& J^3_n\,;\,\psi_n\rightarrow \psi_n\,,\\
&&
J^-_n=i(-\cosh  y \, \del_y +K_n \sinh  y) \,,\\
&&
J^+_{n+1}=i(\cosh y \, \del_y +K_n \sinh y )\,,\\
&& J^3_n \psi_n =K_n \psi_n\,.
\eeq
These currents have been considered in 
solvabe quantum mechanical models in one dimension. 
Here we use the convention in the paper  
\cite{Cevik:2016mnr}. 
But our convention is slightly different from that of the paper.

When the operators act on 
the set of solutions $\{\psi_n\}$, 
one can show the operators
 satisfy commutation relations
that represent $su(1,1)$ algebra
\beq
&& [J^3, J^{\pm}]=\pm J^{\pm}\,,\,\, [J^+ ,J^-]=-2 J^3\,.
\eeq
These are obtained from the following calculation
\beq
&& 
J^3_n J^3_n \psi_n =K_n^2 \psi_n\,,\\
&& J^+_n J^-_n\psi_n =\cosh^2 y \cdot \del^2_y \psi_n 
-K_n 
(1+K_n \sinh^2 y )\cdot \psi_n\,,\\
&& J^-_{n+1} J^+_{n+1}\psi_n =\cosh^2 y \cdot \del^2_y \psi_n 
-K_n 
(-1+K_n \sinh^2 y )\cdot \psi_n\,,\\
&& J^3_{n-1}J^-_n \psi_n =K_{n-1} J^-_n \psi_n\,,\\
&& J^-_{n}J^3_n \psi_n =K_{n} J^-_n \psi_n\,,\\
&& J^3_{n+1}J^+_{n+1} \psi_n =K_{n+1} J^+_{n+1} \psi_n\,,\\
&& J^+_{n+1}J^3_n \psi_n =K_{n} J^+_{n+1} \psi_n\,.
\eeq

Next we rewrite the Schr\"odinger equation
\beq
&& \left[\frac{d^2}{dy^2}-K_n^2 +\frac{\lambda (\lambda -1)}{\cosh^2 y}\right]\psi_n=0
\eeq
into the form
\beq 
&&\cosh^2 y \cdot \del_y^2 \psi_n =K^2_n \cosh^2y \cdot \psi_n 
-\lambda (\lambda -1)\psi_n \,,
\eeq
and evaluate the bilinears of the generators
\beq
&& J^3_n J^3_n \psi_n =K^2_n \psi_n\,,\\
&& 
-J^+_n J^-_n \psi_n -J^-_{n+1}J^+_{n+1}\psi_n \nonumber \\
&&\qquad 
=-2 \cosh^2 y \cdot \del^2_y \cdot \psi_n +2 K_n^2 \sinh^2 y \cdot \psi_n \\
&& \qquad 
=-2[K^2_n \cosh^2y \cdot \psi_n -\lambda (\lambda -1)\psi_n ]
+2 K_n^2 \sinh^2 x \cdot \psi_n \\
&&\qquad =-2 K^2_n \psi_n 
+2\lambda (\lambda -1)\psi_n\,,
\eeq
to obtain
\beq
(-J^+J^- -J^- J^+ +2 J^3 J^3)\psi_n =2\lambda (\lambda -1)\psi_n\,.
\eeq
Then 
one can write down the Casimir invariant $J^2$ in this model by using the generators
\beq
&&
J^2=-J^3J^3 +\frac{1}{2}(J^+J^- +J^-J^+)=-\lambda (\lambda -1)\,.
\eeq
The representation $su(1,1)$ is classified in terms of  the eigenvalue of $J^2$. 
It is related to the height $V_0$ of the peak of $V(x)$
and we can write down the $V_0$ in terms of the Casimir $J^2$
\beq
&& V_0 =-\alpha^2 \lambda (\lambda -1) =\alpha^2 J^2.
\eeq
We make a remark here: the Casimir $J^2$ is the hermitian operator 
and its eigenvalue is real number. But the generator $J^3$ is not hermitian and 
its eigenvalue $K_n$ can take its value in complex number 
$n+\frac{1}{2}\pm i\nu$ $(\nu\in {\bf R})$, namely, 
the representation could be non-unitary.

\bibliographystyle{utphys}
\bibliography{ref}

\end{document}